# Rules of Thumb for Information Acquisition from Large and Redundant Data

Wolfgang Gatterbauer

Computer Science and Engineering
University of Washington, Seattle
gatter@cs.washington.edu

**Abstract.** We develop an abstract model of information acquisition from redundant data. We assume a random sampling process from data which provide information with bias and are interested in the fraction of information we expect to learn as function of (*i*) the sampled fraction (*recall*) and (*ii*) varying bias of information (*redundancy distributions*). We develop two rules of thumb with varying robustness. We first show that, when information bias follows a Zipf distribution, the 80-20 rule or Pareto principle does surprisingly not hold, and we rather *expect to learn less than 40% of the information when randomly sampling 20% of the overall data*. We then analytically prove that for large data sets, randomized sampling from power-law distributions leads to "truncated distributions" with the same power-law exponent. This second rule is very robust and also holds for distributions that deviate substantially from a strict power law. We further give one particular family of power-law functions that remain completely invariant under sampling. Finally, we validate our model with two large Web data sets: link distributions to domains and tag distributions on delicious.com.

## 1 Introduction

The 80-20 rule (also known as Pareto principle) states that, often in life, 20% of effort can roughly achieve 80% of the desired effects. An interesting question is as to weather this rule also holds in the context of information acquisition from redundant data. Intuitively, we know that we can find more information on a given topic by gathering a larger number of data points. However, we also know that the marginal benefit of knowing additional data decreases with the size of the corpus. Does the 80-20 rule hold for information acquisition from redundant data? Can we learn 80% of URLs on the Web by parsing only 20% of the web pages? Can we learn 80% of the used vocabulary by looking at only 20% of the tags? Can we learn 80% of the news by reading 20% of the newspapers? More generally, *can we learn 80% of all available information* in a corpus *by randomly sampling 20% of data* without replacement?

In this paper, we show that when assuming a Zipf redundancy distribution, the Pareto principle does not hold. Instead, we rather expect to see less than 40% of the available information. To show this in a principled, yet abstract fashion, we develop an *analytic sampling model* of information acquisition from redundant data. We assume the *dissemination* of relevant information is biased, i.e. different pieces of information are more or less frequently represented in available sources. We refer to this bias as *redundancy distribution* in accordance with work on redundancy in information extraction [11]. Information acquisition, in turn, can be conceptually broken down into the subsequent steps of IR, IE, and II, i.e. visiting a fraction $r$ of the available sources, extracting the information, and combining it into a unified view (see Fig. 1). Our model relies on only three simple abstractions: (1) we consider a purely *randomized sampling* process without replacement; (2) we do not model *disambiguation* of the data, which is a major topic in information extraction, but not our focus; and (3) we consider the process in the limit of *infinitely large data sets*. With these three assumptions, we estimate the success of information acquisition as function of the (*i*) *recall* of the retrieval process and (*ii*) *bias in redundancy* of the underlying data.

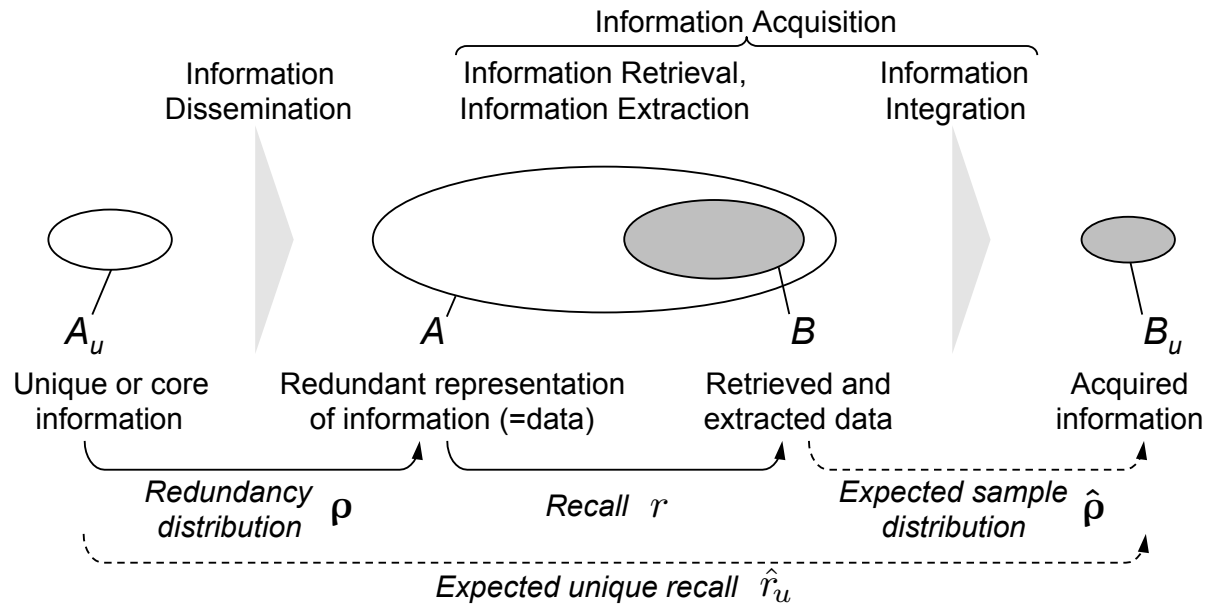

**Fig. 1: Processes of information dissemination and information acquisition. We want to predict the fraction of information we can learn ($\hat{r}_u$) as a function of recall ($r$) and the bias in the data (redundancy distribution ρ).**

**Main contributions.** We develop an analytic model for the information acquisition from redundant data and (1) derive the 40-20 rule, a modification of the Pareto principle which has not been stated before. (2) While power laws do not remain invariant under sampling in general [26], we prove that one particular power law family does remain invariant. (3) While other power laws do not remain invariant in their overall shape, we further prove that the "core" of such a frequency distribution does remain invariant; this observations allows us to develop a second rule of thumb. (4) We validate our predictions by randomly sampling from two very large real-world data sets with power-law behavior.

This is the full version of a conference paper [16] (pages 1–14). All proofs and further details are contained in the appendix.

## 2 Basic notions used throughout this paper

We use the term *redundancy* as synonym for frequency or multiplicity. We do so to remain consistent with the term commonly used in web information extraction, referring to the redundant nature of information on the Web. The notions

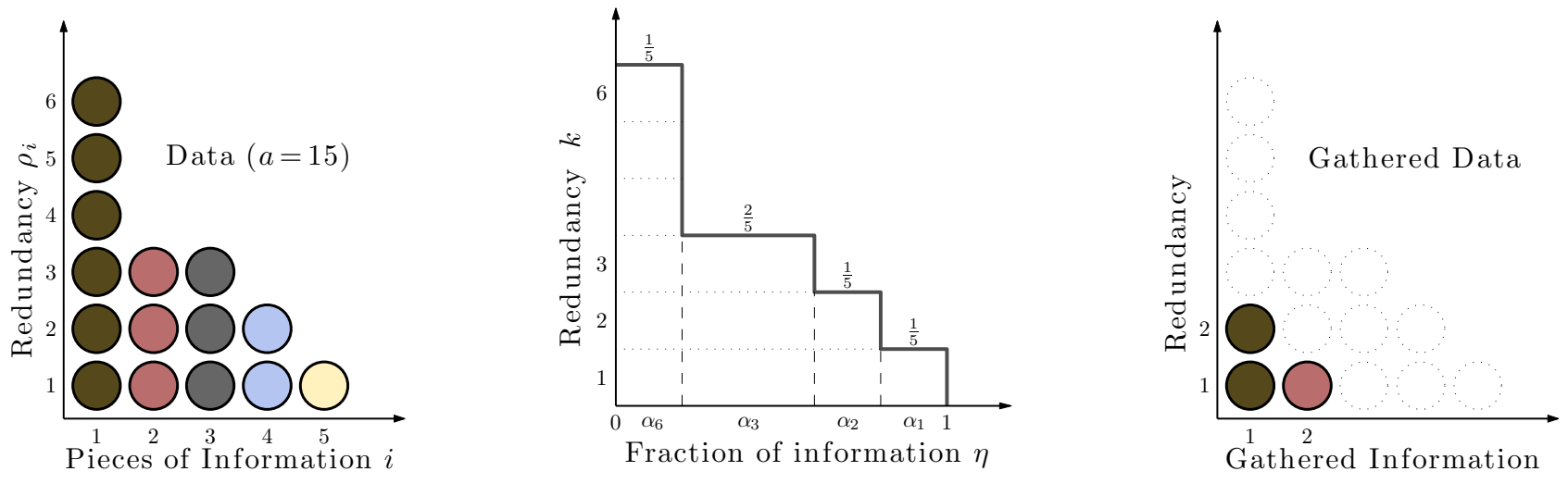

**(a)** $\boldsymbol{\rho}=(6,3,3,2,1)$ **(b)** $\boldsymbol{\alpha}=(\frac{1}{5},\frac{1}{5},\frac{2}{5},0,0,\frac{1}{5})$ **(c)** $r=\frac{3}{15}, r_u=\frac{2}{5}$

**Fig. 2: (a):** ***Redundancy distribution*** $\boldsymbol{\rho}$**. (b): Representation as *redundancy frequency distribution*** $\boldsymbol{\alpha}$**. (c): Recall** $r=\frac{3}{15}$ **and *unique recall*** $r_u=\frac{2}{5}$**.**

of data and information are defined in various and partly contradicting ways in the information retrieval, information extraction, database and data integration literature. In general, their difference is attributed to novelty, relevance, organization, available context or interpretation. The most commonly found understanding is that of *data as representation of information* which can become information when it is interpreted as new and relevant in a given context [6]. In this work, we follow this understanding of data as "raw" information and use the term data for the partly redundant representation of information.

Let $a$ be the total number of data items and $a_u$ the number of unique pieces of information among them. *Average redundancy* $\rho$ is simply their ratio $\rho = \frac{a}{a_u}$. Let $\rho_i$ refer to the redundancy of the $i$-th most frequent piece of information. The *redundancy distribution* $\boldsymbol{\rho}$ (also known as rank-frequency distribution) is the vector $\boldsymbol{\rho} = (\rho_1, \ldots, \rho_{a_u})$. Figure 2a provides the intuition with a simple balls-and-urn model: Here, each color represents a piece of information and each ball represents a data item. As there are 3 red balls, redundancy of the information "color = red" is 3. Next, let $\alpha_k$ be the fraction of information with redundancy equal to $k$, $k \in [k_{\max}]$. A *redundancy frequency distribution* (also known as count-frequency plot) is the vector $\boldsymbol{\alpha} = (\alpha_1, \ldots, \alpha_{k_{\max}})$. It allows us to describe redundancy without regard to the overall number of data items $a$ (see Fig. 2b) and, as we see later, an analytic treatment of sampling for the limit of infinitely large data sets. We further use the term *redundancy layer* (also known as complementary cumulative frequency distribution or ccfd) $\eta_k$ to describe the fraction of information that appears with redundancy $\geq k$: $\eta_k = \sum_{i=k}^{k_{\max}} \alpha_i$. For example, in Fig. 2a, the fraction of information with redundancy at least 3 is $\eta_3 = \alpha_3 + \alpha_6$ $= \frac{2}{5} + \frac{1}{5} = \frac{3}{5}$. Finally, *recall* is the well known measure for the coverage of a data gathering or selection process. Let $b$ be a retrieved subset of the $a$ total data items. Recall is then $r = \frac{b}{a}$.

We define *unique recall* as is its counterpart for unique data items. Thus, it measures the coverage of information. Let $b_u$ be the number of unique pieces of information among $b$, and $a_u$ the number of unique pieces of information among $a$. Unique recall $r_u$ is then $r_u = \frac{b_u}{a_u}$. We illustrate again with the urns model: assume that we randomly gather 3 from the 15 total balls (recall $r = \frac{3}{15}$) and that, thereby, we learn 2 colors out of the 5 total available colors (Fig. 2c). Unique recall is thus $r_u = \frac{2}{5}$ and the redundancy distribution of the sample is $\hat{\boldsymbol{\rho}} = (2, 1)$.

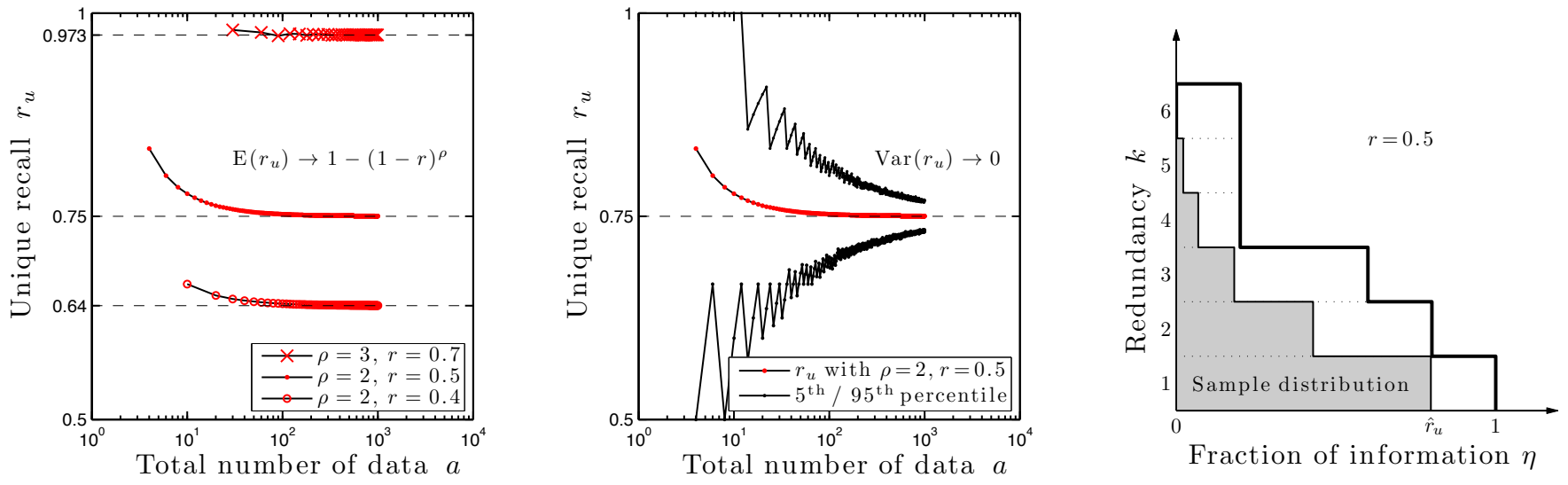

(a) (b) (c)

**Fig. 3: (a, b): Random sampling from an urn filled with $a$ balls in $a_u$ different colors. Each color appears on exactly $\rho = \frac{a}{a_u}$ balls. (c): Normalized sample distribution in grey with unique recall $\hat{r}_u \approx 0.8$ for $r = 0.5$ from $\boldsymbol{\alpha}$ in Fig. 2b.**

## 3 Unique recall

We next give an analytic description of *sampling without replacement* as function of recall and the bias of available information in the limit of very large data sets.

**Proposition 1 (Unique recall $\hat{r}_u$).** *Assume randomized sampling without replacement with recall $r \in [0, 1]$ from a data set with redundancy frequency distribution $\boldsymbol{\alpha}$. The expected value of unique recall for large data sets is asymptotically concentrated around* $\boxed{\hat{r}_u = 1 - \sum_{k=1}^{k_{\max}} \alpha_k (1-r)^k}$.

The proof applies Stirling's formula and a number of analytic transformations to a combinatorial formulation of a balls-and-urn model. The important consequence of Prop. 1 is now that unique recall can be investigated without knowledge of the actual number of data items $a$, but by just analyzing the normalized redundancy distributions. Hence, we can draw general conclusions for families of redundancy distributions assuming very large data sets. To simplify the presentation and to remind us of this limit consideration, we will use the hat symbol and write $\hat{r}_u$ for $\lim_{a \to \infty} \mathbf{E}(r_u) \simeq \mathbf{E}(r_u)$.

Figure 3 illustrates this limit value with two examples. First, assume an urn filled with $a$ balls in $a_u$ different colors. Each color appears on exactly two balls, hence $\rho = 2$ and $a = 2a_u$. Then the expected value of unique recall $r_u$ (fraction of colors sampled) is converging towards $1 - (1-r)^{\rho}$ and its variance towards 0 for increasing numbers of balls $a$ (Fig. 3a, Fig. 3b). For example, keeping $\rho = 2$ and $r = 0.5$ fixed, and varying only $a = 4, 6, 8, 10, \ldots$, then unique recall varies as $r_u = 0.83, 0.80, 0.79, 0.78, \ldots$, and converges towards $\hat{r}_u = 0.75$. At $a = 1000$, $r_u$ is already $0.7503 \pm 0.02$ with 90% confidence. Second, assume that we sample 50% of balls from the distribution $\boldsymbol{\alpha} = (\frac{1}{5}, \frac{1}{5}, \frac{2}{5}, 0, 0, \frac{1}{5})$ of Fig. 2b. Then we can expect to learn $\approx$80% of the colors if $a$ is very large (Fig. 3c). In contrast, exact calculations show that if $a = 15$ as in Fig. 2a, then the actual expectation is around $\approx$79% or $\approx$84% for sampling 7 or 8 balls, respectively. Thus, Prop. 1 calculates the exact asymptotic value only for the limit, but already gives very good approximations for large data sets.

## 4 Unique recall for power law redundancy distributions

Due to their well-known ubiquity, we will next study *power law redundancy distributions.* We distinguish three alternative definitions: (1) power laws in the redundancy distributions, (2) in the redundancy frequencies, and (3) in the redundancy layers. These three power laws are commonly considered to be different expressions of the identical distribution [3,21] because they have the same tail distribution[1], and they are in fact identical in a continuous regime. However, for discrete values, these three definitions of power laws actually produce different distributions and have different unique recall functions. We will show this next.

**Power laws in the redundancy distribution $\boldsymbol{\rho}$.** This distribution arises when the frequency or redundancy $\rho$ of an item is proportional to a power law with exponent $\delta$ of its rank $i$: $\rho(i) \propto i^{-\delta}$, $i \in [a_u]$. Two often cited examples of such power law redundancy distributions where $\delta \approx 1$ are the frequency-rank distribution of words appearing in arbitrary corpora and the size distribution of the biggest cities for most countries. These are called "Zipf Distribution" after [27]. Using Prop. 1 we can derive in a few steps $\hat{r}_{u\rho}(r, \delta) = 1 - \sum_{k=1}^{\infty} \big((2k-1)^{-\frac{1}{\delta}} - (2k+1)^{-\frac{1}{\delta}}\big)(1-r)^k$. For the particularly interesting case of $\delta = 1$, this infinite sum can be reduced to $\hat{r}_{u\rho}(r, \delta{=}1) = \frac{r}{\sqrt{1-r}} \operatorname{artanh}(\sqrt{1-r})$.

**Power laws in the redundancy frequency distribution $\boldsymbol{\alpha}$.** This distribution arises when a fraction of information $\alpha_k$ that appears exactly $k$ times follows a power law $\alpha_k = C \cdot k^{-\beta}$, $k \in \mathbb{N}_1$. Again, using Prop. 1 we can derive in a few steps $\hat{r}_{u\alpha}(r, \beta) = 1 - \frac{\mathrm{Li}_\beta(1-r)}{\zeta(\beta)}$, where $\mathrm{Li}_\beta(x)$ is the polylogarithm $\mathrm{Li}_\beta(x) = \sum_{k=1}^{\infty} k^{-\beta} x^k$, and $\zeta(\beta)$ the Riemann zeta function $\zeta(\beta) = \sum_{k=1}^{\infty} k^{-\beta}$.

**Power laws in the redundancy layers $\boldsymbol{\eta}$.** This distribution arises when the redundancy layers $\eta_k \in [0, 1]$ follow a power law $\eta_k \propto k^{-\gamma}$. From $\eta_1 = 1$, we get $\eta_k = k^{-\gamma}$ and, hence, $\alpha_k = k^{-\gamma} - (k+1)^{-\gamma}$. Using again Prop. 1, we get in a few steps $\hat{r}_{u,\eta}(r, \gamma) = \frac{r}{1-r} \mathrm{Li}_\gamma(1-r)$. For the special case of $\gamma = 1$, we can use the property $\mathrm{Li}_1(x) = -\ln(1-x)$ and simplify to $\hat{r}_{u,\eta}(r, \gamma{=}1) = -\frac{r \ln r}{1-r}$.

**Comparing unique recall for power laws.** All three power laws show the typical *power law tail* in the loglog plot of the redundancy distribution (loglog rank-frequency plot), and it is easily shown that the exponents can be calculated from each other according to Fig. 4e However, the distributions are actually different at the *power law root* (Fig. 4a) and also lead to different unique recall functions. Figure 4b shows their different unique recall functions for the particular power law exponent of $\gamma = 1$ ($\beta = 2$, $\delta = 1$), which is assumed to be

[1] With *tail of a distribution*, we refer to the part of a redundancy distribution for $\eta \to 0$, with *root* to $\eta \to 1$, and with *core* to the interval in between (see Fig. 4a).

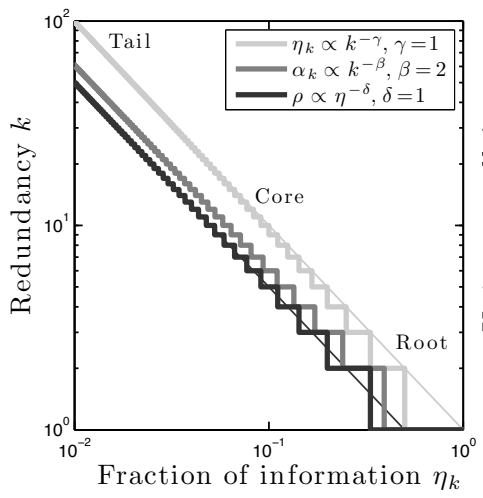

**(a)**

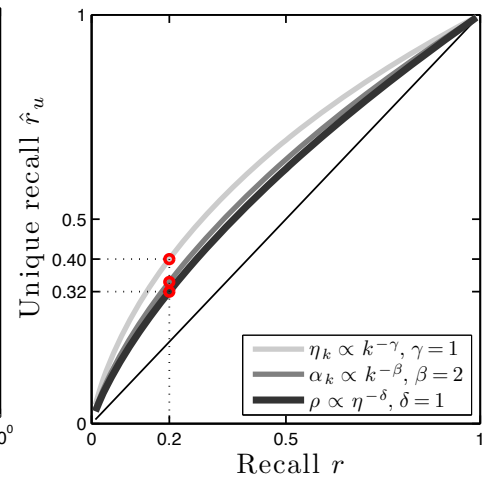

**(b)**

| Assumed power law | Learned fraction |
|---|---|
| $\rho(\eta) \propto \eta^{-1}$ | 32% |
| $\alpha_k \propto k^{-2}$ | 35% |
| $\eta_k \propto k^{-1}$ | 40% |

**(c)**

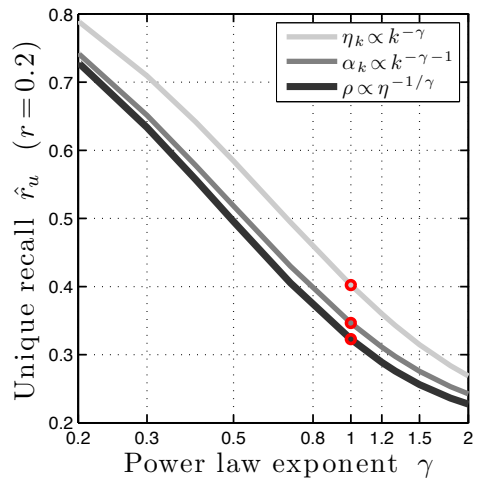

**(d)**

| Power law | Exp | $\delta$ | $\beta$ | $\gamma$ | Defined range | Example values | | | |
|---|---|---|---|---|---|---|---|---|---|
| $\boldsymbol{\rho}: \rho(\eta) \propto \eta^{-\delta}$ | $\delta =$ | – | $\frac{1}{\beta-1}$ | $\frac{1}{\gamma}$ | $0 < \delta < \infty$ | 5 | 2 | 1 | 0.5 |
| $\boldsymbol{\alpha}: \alpha_k \propto k^{-\beta}$ | $\beta =$ | $\frac{1}{\delta}+1$ | – | $\gamma+1$ | $1 < \beta < \infty$ | 1.2 | 1.5 | 2 | 3 |
| $\boldsymbol{\eta}: \eta_k \propto k^{-\gamma}$ | $\gamma =$ | $\frac{1}{\delta}$ | $\beta-1$ | – | $0 < \gamma < \infty$ | 0.2 | 0.5 | 1 | 2 |

**(e)**

**Fig. 4: The three power law redundancy distributions (e) have the same *power law tail* and *power law core* but different *power law roots* in the loglog redundancy plot (a). This leads to different unique recall functions (b&d), and different fractions of information learned after sampling 20% data (c).**

the most common redundancy distribution of words in a large corpus [27] and many other frequency distributions [21,22]. Given our normalized framework, we can now ask the interesting question: *Does the 80-20 rule hold for information acquisition assuming a Zipf distribution?* Put differently, if we sample 20% of the total amount of data (e.g. read 20% of a text corpus, or look at 20% of all existing tags on the Web), what percentage of the contained information (e.g. fraction of different words in a corpus or the tagging data) can we expect to learn if redundancy follows a Zipf distribution? Figure 4c lists the results for the three power law distributions and shows that, depending on which from the three definitions we choose, we can only expect to learn between 32% and 40% of the information. Note that we can apply this rule of thumb without knowing the total amount of available information. Also note that these numbers are sensitive to the power law root and, hence, to deviations from an ideal power law. This is also why unique recall diverges for our 3 variations of power law definitions in the first place (Fig. 4a). Finally, Fig. 4d shows that the power law exponent would have to be considerably different from $\gamma = 1$ to give a 80-20 rule.

**Rule of thumb 1 (40-20 rule).** *When randomly sampling 20% of data whose redundancy distribution follows an exact Zipf distribution, we can expect to learn less than 40% of the contained information.*

## 5 K-recall and the evolution of redundancy distributions

So far, we were interested in the expected fraction $r_u$ of information we learn when we randomly sample a fraction $r$ of the total data. We now generalize the question and derive an analytic description of the overall shape of the *expected sample redundancy distribution* (Fig. 3c). As it turns out, and what will become clear in this and the following section, the natural way to study and solve this question is again to analyze the horizontal "evolution" of the redundancy layers $\boldsymbol{\eta}$ during sampling. To generalize unique recall $r_u$, we define k-recall as the fraction $r_{uk}$ of information that has redundancy $\geq k$ and also appears at least $k$ times in our sample. More formally, let $a_{uk}$ be the number of unique pieces of information with redundancy $\geq k$ in a data set, and let $b_{uk}$ be the number of unique pieces of information with redundancy $\geq k$ in a sample. *K-recall* $r_{uk}$ is then the fraction of $a_{uk}$ that has been sampled: $r_{uk} = \frac{b_{uk}}{a_{uk}}$. The special case $r_{u1}$ is then simply the so far discussed unique recall $r_u$. We assume large data sets throughout this and all following section without always explicitly using the hat notation $\hat{r}_{uk}$.

K-recall has its special relevance when sampling from partly unreliable data. In such circumstances, the general fall-back option is to assume a piece of information to be true when it is independently learned from at least $k$ different sources. This approach is used in statistical polling, in many artificial intelligence applications of learning from unreliable information, and in consensus-driven decision systems: Counting the number of times a piece of information is occurring (its *support*) is used as strong indicator for its truth. As such, *to believe a piece of information only when it appears at least k times in a random sample* serves as starting point from which more complicated polling schemes can be conceived. In this context, $r_{uk}$ gives the ratio of information that we learn *and* consider true (it appears $\geq k$ times in our sample) to the overall information that we would consider true if known to us (it appears $\geq k$ times in the data set) (Fig. 5a).

We also introduce a variable $\omega_k$ for the fraction of total information we get in our sample that appears at least $k$ times instead of just once. Note that $\omega_k = \eta_k r_{uk} = \frac{b_{uk}}{a_u}$. All $\omega_k$ with $k \in [k_{\max}]$ together form the vector $\boldsymbol{\omega}$ representing the *sample redundancy layers* in a random sample with $r \in [0, 1]$. As $r$ increases from 0 to 1, it "evolves" from the $k_{\max}$-dimensional null vector $\mathbf{0}$ to the redundancy layers $\boldsymbol{\eta}$ of the original redundancy distribution. Because of this intuitive interpretation, we call *evolution of redundancy* the transformation of a redundancy frequency distribution given by the redundancy layers $\boldsymbol{\eta}$ to the expected distribution $\boldsymbol{\omega}$ as a function of $r$: $\boldsymbol{\eta} \xrightarrow{r} \boldsymbol{\omega}, r \in [0, 1]$. We further use $\Delta_k$ to describe the fraction of information with redundancy exactly $k$: $\Delta_k = \omega_k - \omega_{k+1}$. To define this equation for all $k \in \mathbb{N}_0$, we make the convention $\omega_0 = 1$ and $\omega_k = 0$ for $k > k_{\max}$. We can then derive the following analytic description:

**Proposition 2 (Sample distribution $\boldsymbol{\omega}$).** *The asymptotic expectation of the fraction of information $\omega_k$ that appears with redundancy $\geq k$ in a randomly sampled fraction $r$ without replacement from a data set with redundancy distribution $\boldsymbol{\alpha}$ is* $\boxed{\hat{\omega}_k = 1 - \sum_{y=0}^{k-1} \sum_{x=y}^{\infty} \alpha_x \binom{x}{y} r^y (1-r)^{x-y}}$ *for* $\lim_{a \to \infty}$.

The first part of the proof constructs a geometric model of sampling from infinitely large data sets with homogenous redundancy and derives the binomial distribution as evolution of the redundancy layers. The second part then applies this result to stratified sampling from arbitrary redundancy distributions.

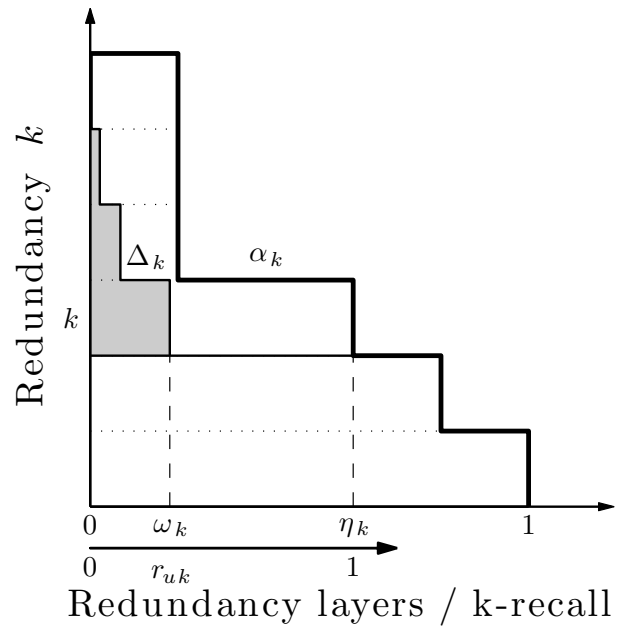

**(a)**

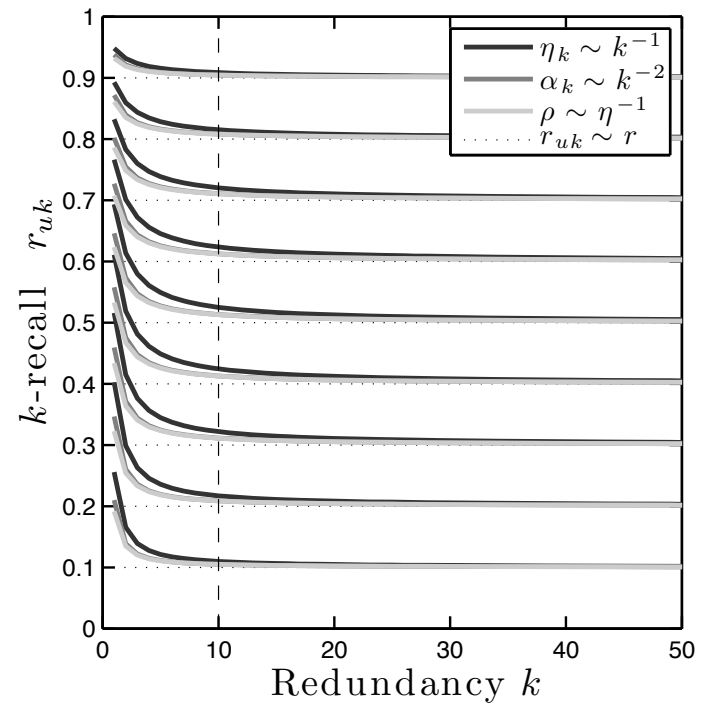

**(b)** $\gamma\!=\!1$

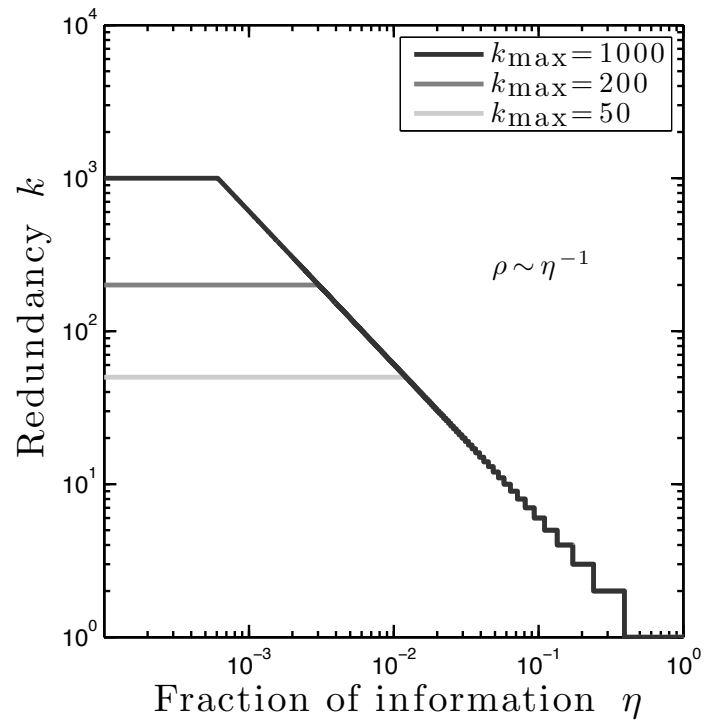

**(c)** $\gamma\!=\!1$

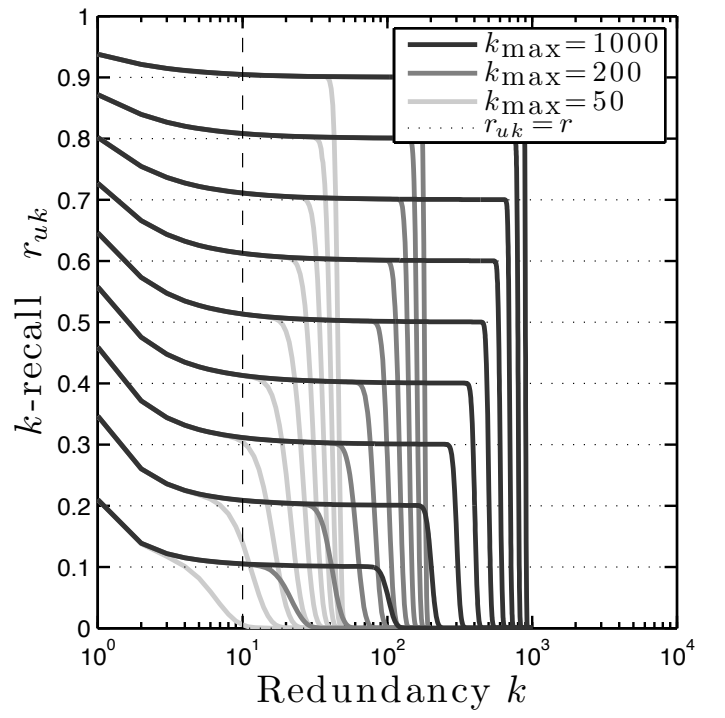

**(d)** $\gamma\!=\!1$

**Fig. 5: Given a redundancy frequency distribution $\boldsymbol{\alpha}$ and recall $r$. K-recall $r_{uk}$ describes the fraction of information appearing $\geq k$ times that also appears $\geq k$ times in our sample: $r_{uk} = \frac{\omega_k}{\eta_k}$ (a). Sampling from *completely developed power laws* leads to sample distributions with the same power law tail, and $r_{uk} \approx r^{\gamma}$ holds independent of $k$ for $k \gtrsim 10$ (b). *Truncated power laws* are cut off at some maximum value $k_{\max}$ (c). As a consequence, the tails of the sample distributions "break in" for increasingly lower recalls (d). However, the invariant power law core with $r_{uk} \approx r^{\gamma}$ is still visible.**

## 6 The Evolution of power laws

Given the complexity of Prop. 2, it seems at first sight that we have not achieved much. As it turns out, however, this equation hides a beautiful simplicity for power laws: namely, their overall shape remains "almost" invariant during sampling. We will first formalize this notion, then prove it, and finally use it for another, very robust rule of thumb.

We say a redundancy distribution $\boldsymbol{\alpha}$ is *invariant under sampling* if, independent of $r$, the expected *normalized sample distribution* $\boldsymbol{\Delta}/\omega_1$ *is the same as the* *original distribution*: $\frac{\Delta_k}{\omega_1} = \alpha_k$. Hence, for an invariant distribution it holds that $\omega_k = \sum_{x=k+1}^{\infty} \Delta_x = \omega_1 \sum_{x=k+1}^{\infty} \alpha_x = \omega_1 \eta_k$, and, hence, $r_{uk}$ is independent of $k$: $r_{uk} = \omega_1$. With this background, we can state the following lemma:

**Lemma 1 (Invariant family).** *The following family of redundancy distributions is invariant under sampling:* $\alpha_k = (-1)^{k-1}\binom{\tau}{k}$*, with* $0 < \tau \leq 1$.

The proof of Lemma 1 succeeds by applying Prop. 2 to the invariant family and deriving $r_{uk} = r^{\tau}$ after application of several binomial identities. Note that the invariant family has a power law tail. We see that by calculating its asymptotic behavior with the help of the asymptotic of the binomial coefficient $\binom{\tau}{k} = \mathcal{O}\left(\frac{1}{k^{1+\tau}}\right)$, as $k \to \infty$, for $\tau \notin \mathbb{N}$. Therefore, we also have $\alpha_k = \mathcal{O}\left(k^{-(1+\tau)}\right)$ for $k \to \infty$. Comparing this equation with the power-law in the redundancy frequency plot, $\alpha_k \propto k^{-\beta}$, we get the power-law equivalent exponent as $\beta = \tau + 1$, with $1 < \beta \leq 2$. Also note that the invariant family is not "*reasonable*" according to the definition of [2], since the mean redundancy $\sum_{k=1}^{\infty} \eta_k$ is not finite.

We next analyze sampling from *completely developed power laws*, i.e. distributions that have infinite layers of redundancy ($k_{\max} \to \infty$). Clearly, those cannot exist in real discrete data sets, but their formal treatment allows us to also consider sampling from *truncated power laws*. The latter are real-world power law distributions which are truncated at $k_{\max} \in \mathbb{N}$ (Fig. 5c). We prove that the power law core remains invariant for truncated power laws, and they, hence, appear as "almost" invariant over a large range, i.e. except for their tail and their root.

**Lemma 2 (Completely developed power laws).** *Randomized sampling without replacement from redundancy distributions with completely developed power law tails* $\boldsymbol{\alpha}_C$ *leads to sample distributions with the same power law tails.*

**Theorem 1 (Truncated power law distributions).** *Randomized sampling without replacement from redundancy distributions with truncated power law tails* $\boldsymbol{\alpha}_T$ *leads to distributions with the same power law core but further truncated power law tails.*

The proof for Lemma 2 succeeds in a number of steps by showing that $\lim_{i,j\to\infty} \frac{r_{ui}}{r_{uj}} = 1$ for distributions with $\boldsymbol{\alpha}_C$. The proof of Theorem 1 builds upon this lemma and shows that $\lim_{k \ll k_{\max}} \Delta_k(\boldsymbol{\alpha}_T, r) = \Delta_k(\boldsymbol{\alpha}_C, r)$. In other words, Theorem 1 states that sampling *from real-world power law distributions* leads to distributions with the same power law core but possibly different tail and root. More formally, $r_{uk} \approx r^{\gamma}$ for $k_1 < k < k_2$, where $k_1$ and $k_2$ depend on the actual distribution, maximum redundancy and the power law exponent. Both, tail and root, are usually ignored when judging whether a distribution follows a power law (cf. Figure 3 in [8]), and to the best of our knowledge, this result is new. Furthermore, it is only recently that Stumpf et al. [26] have shown that sampling from power laws does not lead to power laws in the sample, in general. Our results clarifies this result and shows that *only their tails and roots are subject to change.* Figure 5b, Fig. 5c and Fig. 5d illustrate our result.

**Rule of thumb 2 (Power law cores).** *When randomly sampling from a power law redundancy distribution, we can expect the sample distribution to be a power law with the same power law exponent in the core:* $\boxed{r_{uk} \approx r^{\gamma} \textit{ for } k_1 < k < k_2}$.

## 7 Large real-world data sets

**Data sets.** We use two large real-world data sets that exhibit power-law characteristics to verify and illustrate our rules of thumb: the number of links to web domains and the keyword distributions in social tagging applications.

(1) The first data set is a snapshot of a top level domain in the World Wide Web. It is the result of a complete crawl of the Web and several years old. The set contains 267,415 domains with 5.422,730 links pointing between them. From Fig. 6a, we see that the redundancy distribution follows a power law with exponent $\gamma \approx 0.7$ ($\beta \approx 1.7$, $\delta \approx 1.43$) for $k \gtrsim 100$. Below 100, however, the distribution considerably diverges from this exponent, which is why we expect that rule of thumb 1 does not apply well. We now assume random sampling amongst all links in this data set (e.g. we randomly choose links and discover new domains) and ask: ($i$) what is the expected number of domains and their relative support (as indicated by linking to it) that we learn as function of the percentage of links seen? ($ii$) what is the fraction of domains with support $\geq k$ in the original data that we learn with the same redundancy?

(2) The second data set concerns different keywords and their frequencies used on the social bookmarking web service Delicious (`http://delicious.com`). A total number of $\approx 140$ Mio tags are recorded of which $\approx 2.5$ Mio keywords are distinct [7]. The redundancy distribution (Fig. 6c) follows a power law with exponent $\gamma \approx 1.3$ ($\beta \approx 2.3$, $\delta \approx 1.3$) very well except for the tail and the very root. Here we assume random sampling amongst all individual tags given by users (e.g. we do not have access to the database of Delicious, but rather crawl the website) and ask: ($i$) what is the expected number of different tags and their relative redundancies that we learn as function of the percentage of all tags seen? ($ii$) what is the fraction of important tags in the sample (tags with redundancy at least $k$) that we can also identify as important by sampling a fraction $r$?

**Results.** From Fig. 6b and Fig. 6d we see that after sampling 20% of links and tags respectively, we learn 60% and 40% of the domains and words, respectively. Hence, our first rule of thumb works well only for the second data set which better follows a power law. Our second rule of thumb, however, works well for both data sets: In Fig. 6b, we see that, in accordance with our predication, the horizontal lines for $r_{uk} = r^{\gamma}$ become apparent for $10^2 < k < 10^3$, and in Fig. 6d, for $10^1 < k < 10^4$ (compare with our prediction in Fig. 5d).

## 8 Related Work

Whereas the influence of redundancy of a search process has been widely analyzed [5,19,25], and randomized sampling used in other papers in this field [11,19], our approach is new in the way that we analytically characterize the behavior of the sampling process as a function of $(i)$ the bias in redundancy of the data and $(ii)$ recall of the used retrieval process. In particular, this approach allows us to prove a to date unknown characteristics of power laws during sampling. Achlioptas et al. [2] give a mathematical model that shows that traceroute sampling from Poisson-distributed random graphs leads to power laws. Their analysis is limited to "reasonable" power laws, which are such for which $\alpha > 2$ and also assumes a very concrete sampling process tailored to their context. This is in contrast with our result which proves that completely developed power law functions retain their power law tail, and truncated power laws at least their power law core during sampling. Haas et al. [18] and Chaudhuri et al. [9] investigate ways to estimate the number of different attribute values in a given database. This problem is related in its background but different from its focus. We estimate the number of unique attributes seen after sampling a fraction and later the overall sample distribution. Stump et al. [26] show that, in general, power laws do not remain invariant under sampling. In this paper, we could show that – while not in their entirely – at least the *core of power laws remains invariant under sampling*. General balls-and-urn models have been treated in detail by Gardy [14]. Gardy showed a general theorem which contains Prop. 1 as a special case. However, she does neither investigate the behavior of power laws during sampling, nor extends this result to the evolution of the overall distribution Prop. 2. Only the later allowed us to investigate the overall shape of redundancy distributions during sampling. Flajolet and Sedgewick [12] study the evolution of balanced, single urn models of finite dimensions under random sampling, where dimensionality refers to the number of colors. Using methods of analytic combinatorics, they can associate an ordinary differential system of the same dimension to any balanced urn model, and that an explicit solution of the differential systems provides automatically an analytic solution of the urn model. They mainly focus on urn models of dimension 2 (i.e., balls can be of either of two colors), and also solve some special cases for higher dimensions. They further note, that there is no hope to obtain general solutions for higher dimensions, however, that special cases warrant further investigation. Using a similar, but slightly different nomenclature, we also studied a special case of balanced, single urn models, however with infinite dimension (i.e., infinite number of colors). We further showed that the case of infinite dimensions allows simple analytic solutions which very closely represent cases with high dimensionality.

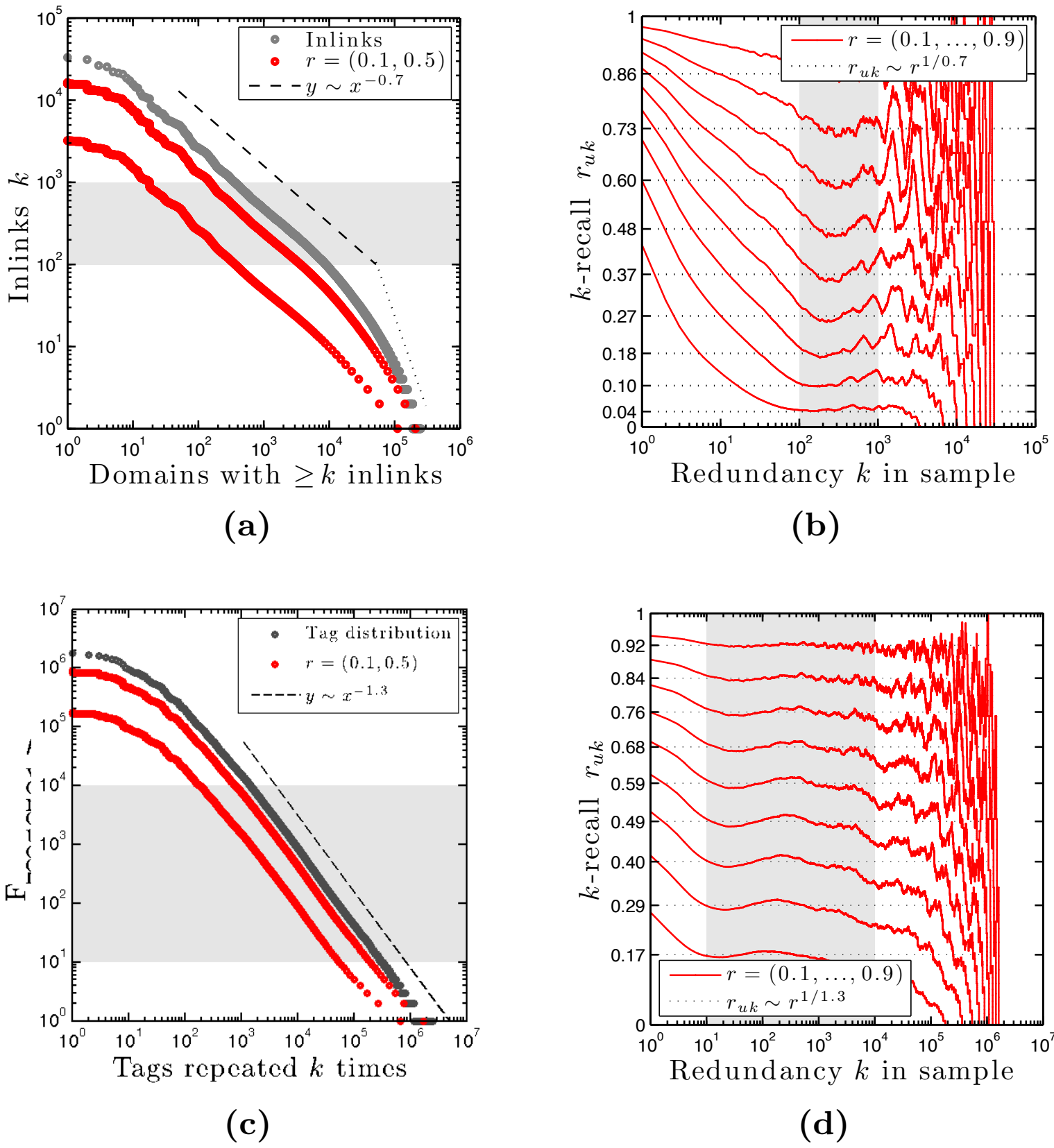

**Fig. 6: Original and sample web link distribution (a). Resulting k-recalls averaged over $N = 100$ repetitions (b). Original and sample tag distribution on Delicious (c). Resulting k-recalls averaged over $N = 10$ repetitions (d).**

In [15], we gave Prop. 1 and motivated the role of different families of redundancy distributions on the effectiveness of information acquisition. However, we did not treat the case of power laws, nor the evolution of distributions during sampling (Prop. 2). To the best of our knowledge, the main results in this paper are new. Our analytic treatment of power laws during sampling, the invariant family, and the proof that sections of power laws remain invariant are not mentioned in any prior work we are aware of (cf. [10,12,13,14,21,22,24]).

## 9 Discussion and Outlook

Our target with this paper was to develop a general model of the information acquisition process (retrieval, extraction and integration) that allows us to estimate the overall success rate when acquiring information from redundant data. With our model, we derive the 40-20 rule of thumb, an adaptation of the Pareto principle. This is a negative result as to what can be achieved, in general. A crucial idea underlying our mathematical treatment of sampling was adopting a horizontal perspective of sampling and thinking in layers of redundancy ("k-recall"). Whereas our approach assumes an infinite amount of data, we have shown our approximation holds very well for large data sets (see Fig. 2b). We have focused on power laws, as they are the dominant form of biased frequency distributions. Whereas Stump et al. [26] have shown that, in general, power laws do not remain invariant under sampling, we have shown that (*i*) there exists one concrete family of power laws which *does* remain invariant, and (*ii*) while power laws do not remain invariant in their tails and root, their core does remain invariant. And we have used this observation to develop a second rule of thumb which turns out to be very robust (cp. Fig. 5d with Fig. 6d). In future work, we intend to extend this analytic method to depart from the pure randomized sampling assumption and incorporate more complicated retrieval processes.

**Acknowledgements.** This work was partially supported by a DOC scholarship from the Austrian Academy of Sciences, by the FIT-IT program of the Austrian Federal Ministry for Transport, Innovation and Technology, and by NSF IIS-0915054. The author would like to thank one of the major search engines for access to the web link data, Ciro Catutto and the TAGora project for access to the Delicious tag data, and Bernhard Gittenberger for mathematical insights and a survey of the related statistical literature. More details will be available on the project page: `http://uniquerecall.com`

# A Nomenclature

| | |
|---|---|
| $a$ | total data items |
| $b$ | sampled data items |
| $r$ | recall or coverage of data $= b/a$ |
| $[\,]_u$ | subscript for "unique" information contained in data |
| $a_u$ | total pieces of information |
| $b_u$ | acquired pieces of information |
| $r_u$ | unique recall or coverage of information $= b_u/a_u$ |
| $\rho$ | average redundancy $= a/a_u$ |
| $\rho_i$ | redundancy of a piece of information with rank $i$ |
| $\boldsymbol{\rho}$ | redundancy distribution $= (\rho_1, \ldots, \rho_{a_u})$ |
| $\hat{[\,]}$ | accent for approximation by limes $= \lim_{a\to\infty} \mathbf{E}([\,])$ |
| $\hat{r}_u$ | approximate unique recall |
| $[\,]_{uk}$ | subscript for information with redundancy $k$ |
| $a_{uk}$ | pieces of information with redundancy $\geq k$ |
| $b_{uk}$ | pieces of information acquired with redundancy $\geq k$ |
| $k_{\max}$ | maximum redundancy |
| $\alpha_k$ | fraction of information with redundancy $k$ |
| $\boldsymbol{\alpha}$ | redundancy frequency distribution $= (\alpha_1, \ldots, \alpha_{k_{\max}})$ |
| $\eta_k$ | fraction of information with redundancy $\geq k$ |
| $\boldsymbol{\eta}$ | normalized redundancy layers $= (\eta_1, \ldots, \eta_{k_{\max}})$ |
| $\omega_k$ | evolution of the $k$-th redundancy layer $= b_{uk}/a_u$ |
| $\boldsymbol{\omega}$ | vector of sample evolution $= (\omega_1, \ldots, \omega_{k_{\max}})$ |
| $r_{uk}$ | $k$-recall $= b_{uk}/a_{uk}$ |
| $\mathbf{r_u}$ | vector of $k$-recall $= (r_{u1}, \ldots, r_{uk_{\max}})$ |
| $\Delta_k$ | fraction of information with redundancy $= k$ in a sample |
| $\boldsymbol{\Delta}$ | redundancy frequency distribution of sample $= (\Delta_1, \ldots, \Delta_{k_{\max}})$ |

**(a)**

| | Original redundancy distribution | Sample redundancy distribution | Relative fraction |
|---|---|---|---|
| Fraction with redundancy $= k$ | $\alpha_k \in \boldsymbol{\alpha}$ | $\Delta_k \in \boldsymbol{\Delta}$ | $\theta_k = \frac{\Delta_k}{\alpha_k}$ |
| Fraction with redundancy $\geq k$ | $\eta_k \in \boldsymbol{\eta}$ | $\omega_k \in \boldsymbol{\omega}$ | $r_{uk} = \frac{\omega_k}{\eta_k}$ |

**(b)**

**Fig. 7: Variables used in this paper (a). Original and sample redundancy distributions and their ratios (b). Also compare with Fig. 13a and Fig. 13b.**

One basic abstraction used throughout this paper is that of a redundancy distribution, illustrated with a colored balls and urn model in Fig. 2a. The vertical axis shows redundancy for each color and the horizontal axis lists colors in order of decreasing values. The two axes *information* and *redundancy* span the area of *data*. In short: $\boxed{\text{data} = \text{information} \times \text{redundancy}}$.

# B Details Section 3 (unique recall)

## B.1 Unique recall for uniform distributions

Here, we first show Prop. 1 for the special case of a uniform redundancy distribution $\rho = \text{const}$, then treat the general case in Section B.2.

**Lemma 3 (Uniform unique recall).** *Assume randomized sampling from a data set with uniform redundancy distribution* $\boldsymbol{\rho}$ *with* $\rho_i = \rho$*, and let* $r$ *be the recall of the underlying data gathering process. Then the asymptotic expectation of unique recall* $r_u$ *for large data sets is*

$$\lim_{a\to\infty} \mathbf{E}\left(r_u\right) = \boxed{\hat{r}_u = 1 - (1-r)^\rho} \ . \tag{1}$$

*Proof.* Assume an urn filled with balls in $a_u$ different colors (Fig. 8). Each color appears on exactly $\rho$ different balls, which makes a total number of $a = \rho a_u$ balls. We now randomly draw $b$ balls from the urn without replacement. What is the average number of different colors $b_u$ we are expected to see?

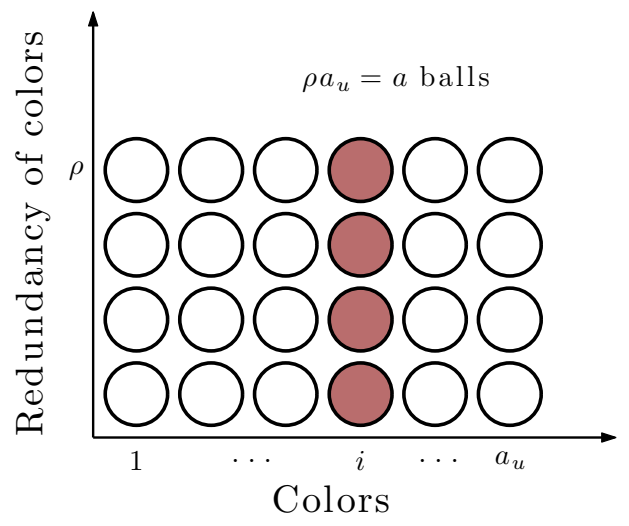

**Fig. 8: An urn filled with $a$ balls in $a_u$ different colors.**

When randomly drawing $b$ from $a$ balls, the outcome is any of $\binom{a}{b}$ equally likely subsets. The number of those subsets in which any given color $i$ does not appear is the number of possible subsets when choosing $b$ from $a-\rho$ balls, $\binom{a-\rho}{b}$. Hence, the likelihood that any color $i$ does not appear in a random sample is the fraction of those numbers, or

$$\mathbf{P}\left[X(i)=0\right] = \frac{\#\text{ of subsets without } i}{\#\text{ of total subsets}} = \frac{\binom{a-\rho}{b}}{\binom{a}{b}} \ .$$

The likelihood of color $i$ appearing in the sample at least once is the compliment of this fraction, or

$$\mathbf{P}\left[X(i)\geq 1\right] = 1 - \frac{\binom{a-\rho}{b}}{\binom{a}{b}} \ . \tag{2}$$

As this equation holds for each color independently and we can, therefore, treat the likelihood of appearance for any color as independent events, the expected number of different colors $b_u$ in the draw is equal to

$$\mathbf{E}\left(b_u\right) = a_u \mathbf{P}\left[X(i) \geq 1\right] = a_u \left(1 - \frac{\binom{a-\rho}{b}}{\binom{a}{b}}\right) .$$

Therefore, the expected value of unique recall is

$$\mathbf{E}\left(r_u\right) = \frac{\mathbf{E}\left(b_u\right)}{a_u} = 1 - \frac{\binom{a-\rho}{b}}{\binom{a}{b}}. \tag{3}$$

Now, let us compute the asymptotic expectation for $a \to \infty$ and $b$ proportional to $a$. We have

$$\frac{\binom{a-\rho}{b}}{\binom{a}{b}} = \begin{cases} \frac{(a-\rho)!}{a!} \frac{(a-b)!}{(a-b-\rho)!} & \text{if } b < a - \rho \\ 0 & \text{if } b \geq a - \rho. \end{cases} \tag{4}$$

Now note that $a$ and $a - b$ tend to infinity whereas $\rho$ is constant. Thus we have to analyze expressions like $\frac{(a-\rho)!}{a!}$ asymptotically (then plug $a - b$ instead of $a$ into this formula to get the asymptotic equivalent for the second factor of Eq. 4). This can be done by Stirling's formula

$$n! = \frac{n^n}{e^n}\sqrt{2\pi n}\left(1 + \frac{1}{12n} + \mathcal{O}\left(\frac{1}{n^2}\right)\right), \text{ as } n \to \infty.$$

In a few steps we can derive

$$\begin{aligned} \lim_{a\to\infty} \frac{\binom{a-\rho}{b}}{\binom{a}{b}} &= \lim_{a\to\infty} \frac{(a-\rho)!}{a!} \frac{(a-b)!}{(a-b-\rho)!} \\ &= \lim_{a\to\infty} \frac{(a-b)^\rho}{a^\rho} = (1-r)^\rho. \end{aligned} \tag{5}$$

So we have finally

$$\lim_{a\to\infty} \mathbf{E}\left(r_u\right) = \hat{r}_u = 1 - (1-r)^\rho. \qquad \square$$

## B.2 Unique recall for general distributions

The previous approach of deriving the basic unique recall formula allows us to treat general redundancy distributions as well. Only now, each color $i$ appears with redundancy $\rho(i)$ or $\rho_i$ (Fig. 9).

*Proof (Prop. 1).* From Eq. 2, we know that the likelihood of color $i$ appearing in as random sample of size $b$ at least once is

$$\mathbf{P}\left[X(i) \geq 1\right] = 1 - \frac{\binom{a-\rho_i}{b}}{\binom{a}{b}} .$$

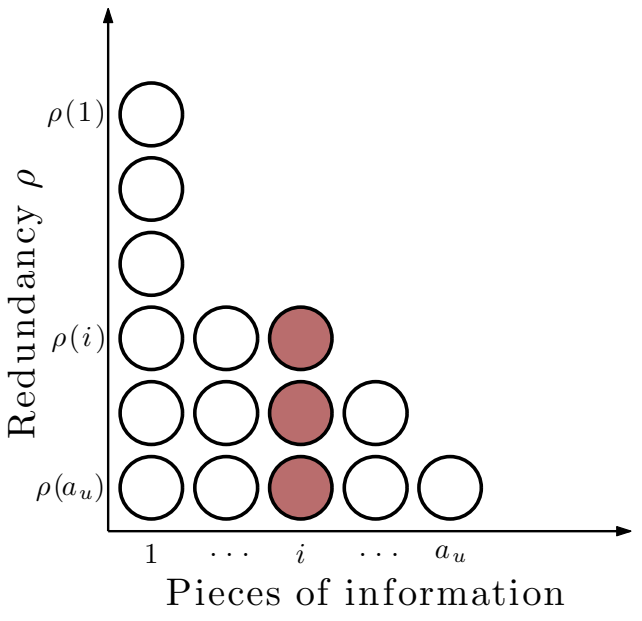

**Fig. 9: An urn filled with balls of $a_u$ different colors with varying redundancy $\rho(i)$ for each color $i$.**

This equation again holds independently for each color $i \in [a_u]$, which allows us to simply add the likelihoods of all colors and calculate the expected number of different colors $b_u$ in the draw as

$$\begin{aligned}\mathbf{E}\left(b_u\right) &= \sum_{i=1}^{a_u} \mathbf{P}\left[X(i) \geq 1\right] \\ &= a_u - \sum_{i=1}^{a_u} \frac{\binom{a-\rho_i}{b}}{\binom{a}{b}} \ .\end{aligned}$$

Therefore, we get as the exact combinatorial expected value $r_u$

$$\begin{aligned}\mathbf{E}\left(r_u\right) &= \frac{\mathbf{E}\left(b_u\right)}{a_u} \\ &= 1 - \frac{1}{a_u} \sum_{i=1}^{a_u} \frac{\binom{a-\rho_i}{b}}{\binom{a}{b}} \\ &= 1 - \frac{1}{a_u \binom{a}{b}} \sum_{i=1}^{a_u} \binom{a-\rho_i}{b} \ . \end{aligned} \tag{6}$$

As we know from our previous limit consideration (Eq. 5)

$$\lim_{a \to \infty} \frac{\binom{a-\rho_i}{b}}{\binom{a}{b}} = (1-r)^{\rho_i} \ ,$$

the exact equation can be simplified for large data sets ($a \to \infty$) to

$$\lim_{a \to \infty} \mathbf{E}\left(r_u\right) = 1 - \frac{1}{a_u} \sum_{i=1}^{a_u} (1-r)^{\rho_i} \ . \tag{7}$$

Whereas the latter formula is much simpler to evaluate for a given redundancy distribution, the bias of redundancy is still described by the exact distribution of individual data items $\boldsymbol{\rho} = (\rho_1, \rho_2, ..., \rho_{a_u})$. However, in a new step we can transform it further to

$$\lim_{a\to\infty} \mathbf{E}\left(r_u\right) = \boxed{\hat{r}_u = 1 - \sum_{k=1}^{k_{\max}} \alpha_k \left(1-r\right)^k}, \tag{8}$$

with $\alpha_k$ standing for the fraction of information with redundancy $k$, $k_{\max}$ being the maximum occurring redundancy $\rho_1$, and the sum of the fractions $\sum_{k=1}^{k_{\max}} \alpha_k = 1$ summing up to 1.

The variance $\mathbf{Var}\left(b_u\right) = \mathbf{E}\left(b_u^2\right) - \left(\mathbf{E}\left(b_u\right)\right)^2$ can be calculated by similar but a bit more intricate calculations as

$$\mathbf{Var}\left(b_u\right) = a_u^2 \left( \frac{\binom{a-2\rho}{b}}{\binom{a}{b}} - \frac{\binom{a-\rho}{b}^2}{\binom{a}{b}^2} \right) + a_u \left( \frac{\binom{a-\rho}{b}}{\binom{a}{b}} - \frac{\binom{a-2\rho}{b}}{\binom{a}{b}} \right)$$

From that, it can be shown that

$$\mathbf{Var}\left(r_u\right) \sim \frac{1}{a_u}(1-r)^\rho \left(1 - (1-r)^\rho \left(1 + \frac{r\rho}{1-r}\right)\right) \tag{9}$$

which tends to 0, as $a \to \infty$. This means that the random variable $r_u$ is asymptotically concentrated around its mean value. □

Note, we can calculate the vector $\boldsymbol{\alpha} = (\alpha_k)$ from $\boldsymbol{\rho}$ as

$$\alpha_k = \frac{\left|\{i|\rho(i) = k, i \in \mathbb{N}_1^{a_u}, \boldsymbol{\rho} = (\rho_i)\}\right|}{a_u} \ ,$$

with $k \in \mathbb{N}_1^{k_{\max}}$, $k_{\max} = \rho_1 = \max(\boldsymbol{\rho})$, and $a_u = \dim(\boldsymbol{\rho})$ being the number of different data items. The vector $\boldsymbol{\alpha}$ presents an *alternative description* of bias in redundancy of data (Fig. 2). However, it is not an *equivalent description* without $a_u$ explicitly stated, which we see by calculating $\boldsymbol{\rho}$ back from $\boldsymbol{\alpha}$ by

$$\rho_i = \min\left\{k \,\middle|\, \sum_{x=k}^{k_{\max}} \alpha_x \geq \frac{i}{a_u}, k \in \mathbb{N}_1^{k_{\max}}, \boldsymbol{\alpha} = (\alpha_x)\right\} \ , \tag{10}$$

with $i \in \mathbb{N}_1^{a_u}$, $k_{\max} = \dim(\boldsymbol{\alpha})$, and the number of total pieces of information $a_u$ not explicitly given by $\boldsymbol{\alpha}$. More formally, we can state for two variations of the mapping:

$$f : \boldsymbol{\rho} \to \boldsymbol{\alpha} : \text{non-injective mapping}$$
$$f : \boldsymbol{\rho} \to (\boldsymbol{\alpha}, a_u) : \text{injective mapping} \ .$$

### B.3 Further observations

**Illustration of the limit value.** Figure 3a illustrates with three example values that the basic unique recall formula poses a good approximation for Eq. 3 as exact combinatorial solution of $\mathbf{E}(r_u)$ for $a \to \infty$ with $a$ being the total number of redundant data items. Figure 3b in addition plots the 5th and 95th percentile of the random variable $r_u$. For this plot, we randomly sampled and averaged over 1000 times for each data point. Given any pair of values $\rho$ and $r$, only certain combinations of $a, b, a_u$ and $b_u$, and thus, certain values of $r_u$ are possible. As a consequence, the resulting percentile graphs are ragged. Finally,

Figure 10 illustrates with our running example Fig. 2 $\hat{r}_u$ is a good approximation not only for $\mathbf{E}(r)$, but also for $\mathbf{E}(b) = a_u\mathbf{E}(r)$. Even the absolute error in $\Delta = \mathbf{E}(b_u) - a_u\hat{r}_u$ generally decreases with the size of the data set.

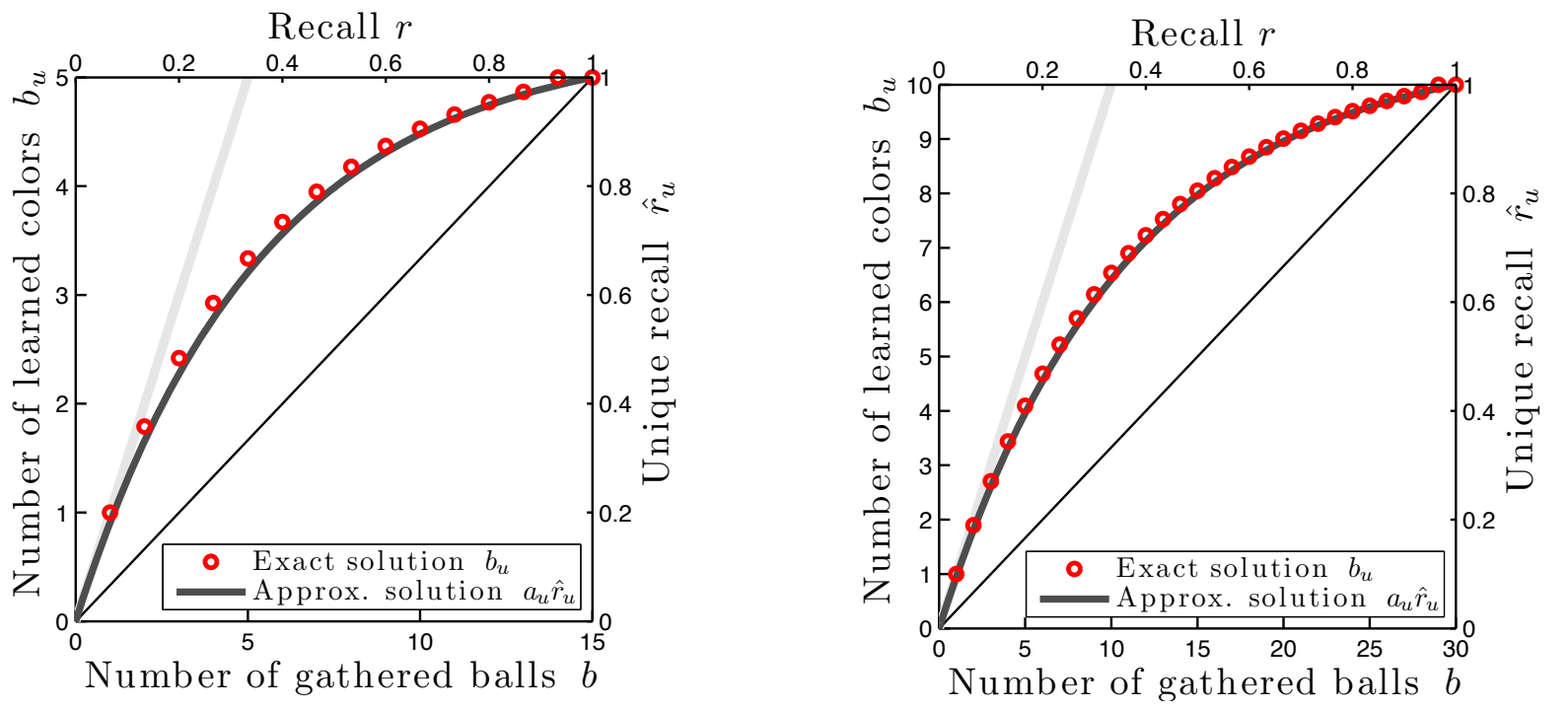

**(a)** $\boldsymbol{\rho}_1$ **(b)** $\boldsymbol{\rho}_2$

| $\boldsymbol{\alpha}$ $a \to \infty$ | | $\boldsymbol{\rho}_1 = (6,3,3,2,1)$ $a_u = 5, a = 15$ | | | | $\boldsymbol{\rho}_2 = (6,6,3,3,3,3,2,2,1,1)$ $a_u = 10, a = 30$ | | | |
|---|---|---|---|---|---|---|---|---|---|
| $r$ | $\hat{r}_u$ | $b$ | $\mathbf{E}(b_u)$ | $a_u\hat{r}_u$ | $\Delta$ | $b$ | $\mathbf{E}(b_u)$ | $a_u\hat{r}_u$ | $\Delta$ |
| 0.2 | 0.455 | 3 | 2.420 | 2.274 | 0.146 | 6 | 4.684 | 4.548 | 0.136 |
| 0.4 | 0.712 | 6 | 3.671 | 3.561 | 0.110 | 12 | 7.230 | 7.123 | 0.107 |
| 0.6 | 0.862 | 9 | 4.369 | 4.308 | 0.061 | 18 | 8.677 | 8.616 | 0.061 |
| 0.8 | 0.949 | 12 | 4.771 | 4.744 | 0.027 | 24 | 9.511 | 9.488 | 0.023 |

**(c)**

**Fig. 10: Comparing $\mathbf{E}(b_u(\boldsymbol{\rho}, r))$ – the exact solution for the expected number of pieces of information learned – for $\boldsymbol{\rho}_1$ and $\boldsymbol{\rho}_2$ with the approximate solution $\mathbf{E}(b_u) \simeq a_u\hat{r}_u(\boldsymbol{\alpha}, r)$ shows that Prop. 1 is a good approximation of the exact solution Eq. 6 for large data sets ($a \to \infty$): For constant $\boldsymbol{\alpha} = (\frac{1}{5}, \frac{1}{5}, \frac{2}{5}, 0, 0, \frac{1}{5})$, not only the error of expected unique recall $\mathbf{E}(r_u) - \hat{r}_u$, but also the error in the expected number of learned pieces of information $\Delta = \mathbf{E}(b_u) - a_u\hat{r}_u$ decreases, in general, with increasing $a$.**

**Analogy.** On a side note, the following problem presents an interesting mathematical analogy: Assume that a web crawler finds each available online copy of a research paper with probability $\gamma$. The probability of missing a document is $1-\gamma$, the probability of missing a document with $c$ copies online $(1-\gamma)^c$, and, hence, the probability of finding and indexing a document with $c$ copies online $1-(1-\gamma)^c$ [23]. Though the nature of the solution is the same as the one to our problem and both problems seem to be identical at first sight, the underlying question is different from asking how many different documents one could retrieve on average. The reason is that expectations cannot be added in the presence of mutual correlations. Looking at one particular document has an exact solution which is always true: $1-(1-\gamma)^c$. The exact answer to our problem is Eq. 3, which only approaches Eq. 1 when taking the limes.

The difference is best illustrated with the first few red dots in Fig. 3a: Expected unique recall is 0.83 for $a=4$ and the $\rho=2$, $r=0.5$ and $a_u=2$ different documents, and not 0.75.

**Geometric interpretation.** Equation 8 also allows an interesting geometric interpretation of unique recall for a general redundancy distribution as the *mean* of all unique recalls $\hat{r}_u(k,r)$ for uniform redundancies $k \leq k_{\max}$, weighted by their fractions $\alpha_k$:

$$\hat{r}_u = \sum_{k=1}^{k_{\max}} \alpha_k \, \hat{r}_u(k,r) \ .$$

The intuition why this formula must hold is the same why *stratification* in statistics does not change the expected outcome of a sampling process. In stratified sampling, first, a population to be sampled is grouped into MECE (mutually exclusive, collectively exhaustive) subgroups and then a fraction is sampled from each strata that is proportional to their relative sizes [24]. The mathematical justification is that, on average, the fractions sampled in a random draw are the same across all strata and the total population. For the same reason, when sampling a fraction $r$ of the total amount of data, $r$ will also be the expected fraction that is sampled from each subset or strata with constant redundancy. Hence, we get back Eq. 8 for the formula of unique recall $\hat{r}_u(\boldsymbol{\alpha},r)$ of a general redundancy distribution $\boldsymbol{\alpha}$:

$$\begin{aligned}
\hat{r}_u(\boldsymbol{\alpha},r) &= \sum_{k=1}^{k_{\max}} \alpha_k r_u(k,r) \\
&= \sum_{k=1}^{k_{\max}} \alpha_k \left(1-(1-r)^k\right) \\
&= 1-\sum_{k=1}^{k_{\max}} \alpha_k (1-r)^k \ .
\end{aligned}$$

We will use this possibility to average over fractions with constant redundancy again in Section D when we calculate the evolution of the general redundancy distribution.

# C Details Section 4 (unique recall for power laws)

## C.1 Power laws in the redundancy distribution

Here, we consider the case when the frequency or redundancy $\rho$ of an item is proportional to a power law with exponent $\delta$ of its rank $i$:

$$\rho(i) \propto i^{-\delta} \ , \qquad i \in [a_u] \ .$$

One often cited example of this distribution with $\delta \approx 1$ (Zipf distribution) is the frequency rank distribution of words appearing in arbitrary corpora [27].

In the normalized redundancy distribution, this power law translates into

$$\rho(\eta) \propto \eta^{-\delta} \ , \qquad \eta \in [0,1] \ ,$$

whereby the above two relations could only hold closely if real values were possible for $\rho(i)$ and $\rho(\eta)$. If we assume some underlying continuous process that is responsible for the observed discrete power law, then the natural way to model above relation is by rounding to the nearest possible redundancy $k \in \mathbb{N}_0$,

$$\begin{aligned} k(\eta) &= \text{round}\big(\rho(\eta)\big) \\ &= \text{round}\big(C \cdot \eta^{-\delta}\big) \\ &= \lfloor C \cdot \eta^{-\delta} + 0.5 \rfloor \ . \end{aligned}$$

The last step uses the floor function $\lfloor x \rfloor$ to describe the greatest integer less or equal to $x$, which, in the next step, helps us calculate the fraction of information $\eta_k$ that appears at least $k$ times. As we only consider positive integers for $k$, we can leave away the floor function when expressing $\eta_k = \eta(k)$ and get

$$k = C \cdot \eta_k^{-\delta} + 0.5$$

or

$$\eta_k = \left(\frac{k - 0.5}{C}\right)^{-\frac{1}{\delta}} \ .$$

As the fraction of information that appears at least one time is equal to 1 and, thus $\eta_1 = 1$, we have $C = 0.5$. So we get

$$\eta_k = (2k-1)^{-\frac{1}{\delta}} \ .$$

From their definitions in Section 2, we know that $\alpha_k = \eta_k - \eta_{k+1}$ and, hence,

$$\alpha_k = (2k-1)^{-\frac{1}{\delta}} - (2k+1)^{-\frac{1}{\delta}} \ . \tag{11}$$

Then, from Prop. 1 we know $\hat{r}_u(r) = 1 - \sum_{k=1}^{\infty} \alpha_k (1-r)^k$ and can now state approximate unique recall for a redundancy distribution that follows a power law with exponent $\delta$ as

$$\hat{r}_u(r) = 1 - \sum_{k=1}^{\infty} \left( (2k-1)^{-\frac{1}{\delta}} - (2k+1)^{-\frac{1}{\delta}} \right) (1-r)^k \ .$$

This infinite sum cannot be reduced in general. However, it does have a closed solution for $\delta = 1$. To see this, we substitute with $x = \sqrt{1-r}$ to get

$$\begin{aligned}
\hat{r}_u(x) &= 1 - \sum_{k=1}^{\infty} \left( (2k-1)^{-\frac{1}{\delta}} - (2k+1)^{-\frac{1}{\delta}} \right) x^{2k} \\
&= 1 - x \sum_{k=1}^{\infty} (2k-1)^{-\frac{1}{\delta}} x^{2k-1} + \frac{1}{x} \sum_{k=1}^{\infty} (2k+1)^{-\frac{1}{\delta}} x^{2k+1} \\
&= 1 - x \sum_{k=1}^{\infty} (2k-1)^{-\frac{1}{\delta}} x^{2k-1} + \frac{1}{x} \sum_{k=0}^{\infty} (2k+1)^{-\frac{1}{\delta}} x^{2k+1} - 1 \\
&= 1 - x \sum_{k=1}^{\infty} (2k-1)^{-\frac{1}{\delta}} x^{2k-1} + \frac{1}{x} \sum_{k=1}^{\infty} (2k-1)^{-\frac{1}{\delta}} x^{2k-1} - 1 \\
&= \frac{1-x^2}{x} \underbrace{\sum_{k=1}^{\infty} (2k-1)^{-\frac{1}{\delta}} x^{2k-1}}_{S} \ .
\end{aligned}$$

It is interesting to observe the relation of the power series $S$ to the polylogarithm: $S$ consists just of the odd terms from the power series of the polylogarithm. To the best knowledge of the author, there is no generally known function defined for this series nor a way to reformulate it as a function of other basic and generally known functions defined in mathematics. For $\delta = 1$, however, $S$ is simply the power series of $\tanh^{-1}(x) = \text{artanh}(x)$, the inverse hyperbolic tangent [4, p. 484] as

$$\text{artanh}(x) = x + \frac{x^3}{3} + \frac{x^5}{5} + \cdots \ .$$

Thus, analogous to the polylogarithm being a generalization of the logarithm with $\text{Li}_1(x) = \log(x)$, $S$ could be considered a similar extension to $\text{artanh}(x)$. Not being defined as such and, therefore, most likely not commonly found in mathematics, we can reformulate unique recall at least for $\delta = 1$ as

$$\hat{r}_u(x) = \frac{1-x^2}{x} \text{artanh}(x) \ .$$

Resubstituting $\sqrt{1-r}$ for $x$, we finally get

$$\hat{r}_u(r) = \frac{r}{\sqrt{1-r}} \text{artanh}(\sqrt{1-r}) \ .$$

### C.2 Power laws in the redundancy frequency plot

Next, we assume that the fraction of information $\alpha_k$ that appears exactly $k$ times follows a power law

$$\alpha_k = C \cdot k^{-\beta} \ , \qquad k \in \mathbb{N}_1 \ .$$

Here, we use $\beta$ as exponent when $\alpha_k$ follows a power law to distinguish this case from the previous one where $\rho(\eta)$ followed a power law with exponent $\delta$. From $k = 1$, we see that the constant of proportionality $C$ is equal to $\alpha_1$ and, hence,

$$\alpha_k = \alpha_1 k^{-\beta} \ .$$

Using the normalizing condition $\sum_{k=1}^{\infty} \alpha_k = 1$, we get

$$\alpha_1 \sum_{k=1}^{\infty} k^{-\beta} = 1 \ .$$

The infinite sum on the left side is known in mathematics as the Riemann zeta function [17, p. 263],

$$\zeta(z) = \sum_{k=1}^{\infty} k^{-z} \ ,$$

which allows to state

$$\alpha_1 = \frac{1}{\zeta(\beta)} \ ,$$

and further

$$\alpha_k = \frac{k^{-\beta}}{\zeta(\beta)} \ .$$

From their definitions in Section 2, we know that the fraction of information $\eta_k$ that appears at least $k$ times (or has redundancy $\geq k$) is

$$\begin{aligned} \eta_k &= 1 - \sum_{x=1}^{k-1} \alpha_k \\ &= 1 - \frac{1}{\zeta(\beta)} \sum_{x=1}^{k-1} x^{-\beta} \ . \end{aligned}$$

The series on the right side is known as the generalized harmonic number of order $(k-1)$ of $\beta$. The generalized harmonic number $H_k^{(z)}$ of order $k$ of $x$ [20, p.74] is defined as

$$H_k^{(z)} = \sum_{x=1}^{k} x^{-z} \ ,$$

which, for $k = \infty$, is equal to the Riemann zeta function:

$$H_\infty^{(z)} = \sum_{x=1}^{\infty} x^{-z} = \zeta(z) \ .$$

We, therefore, have

$$\eta_k = 1 - \frac{H_{k-1}^{(\beta)}}{\zeta(\beta)} \ .$$

Then, from Prop. 1 we know $\hat{r}_u(r) = 1 - \sum_{k=1}^{\infty} \alpha_k (1-r)^k$ and get as approximate unique recall

$$\hat{r}_u = 1 - \frac{1}{\zeta(\beta)} \sum_{k=1}^{\infty} k^{-\beta} (1-r)^k.$$

The infinite series on the right side is known in mathematics as polylogarithm. The polylogarithm $\mathrm{Li}_z(x)$ is defined as

$$\mathrm{Li}_z(x) = \sum_{k=1}^{\infty} k^{-z} x^k \ ,$$

which, for $x = 1$, is again equal to the Riemann zeta function:

$$\mathrm{Li}_z(1) = \sum_{k=1}^{\infty} k^{-z} = \zeta(z) \ .$$

We can, therefore, write unique recall for a redundancy distribution where the redundancy frequencies $\alpha_k$ follow a power law with exponent $\beta$ as

$$\hat{r}_u = 1 - \frac{\mathrm{Li}_\beta(1-r)}{\zeta(\beta)} \ .$$

### C.3 Power laws in the redundancy layers

Here, we assume that the redundancy layers $\eta_k \in [0, 1]$ follow a power law

$$\eta_k \propto k^{-\gamma} \ .$$

As $\eta_1 = 1$ (the first layer must always be 1), we can directly write

$$\eta_k = k^{-\gamma} \ .$$

From $\alpha_k = \eta_k - \eta_{k+1}$ we know the fraction of information that appears exactly with redundancy $k$ to be

$$\alpha_k = k^{-\gamma} - (k+1)^{-\gamma} \ , \tag{12}$$

and from $\hat{r}_u(r) = 1 - \sum_{k=1}^{\infty} \alpha_k (1-r)^k$, we get approximate unique recall as

$$\hat{r}_u(r) = 1 - \sum_{k=1}^{\infty} \left(k^{-\gamma} - (k+1)^{-\gamma}\right) (1-r)^k \ .$$

Like in the previous subsection, the infinite series can be expressed by the polylogarithm; this time, however, after some transformations:

$$\begin{aligned}
\hat{r}_u(r) &= 1 - \sum_{k=1}^{\infty} k^{-\gamma}(1-r)^k + \sum_{k=1}^{\infty}(k+1)^{-\gamma}(1-r)^k \\
&= 1 - \sum_{k=1}^{\infty} k^{-\gamma}(1-r)^k + \frac{1}{1-r}\sum_{k=1}^{\infty}(k+1)^{-\gamma}(1-r)^{k+1} \\
&= 1 - \sum_{k=1}^{\infty} k^{-\gamma}(1-r)^k + \frac{1}{1-r}\sum_{k=2}^{\infty} k^{-\gamma}(1-r)^k \\
&= 1 - \sum_{k=1}^{\infty} k^{-\gamma}(1-r)^k + \frac{1}{1-r}\sum_{k=1}^{\infty} k^{-\gamma}(1-r)^k - 1 \\
&= \frac{r}{1-r}\sum_{k=1}^{\infty} k^{-\gamma}(1-r)^k \ .
\end{aligned}$$

Using the definition for the polylogarithm, we learn unique recall for a redundancy distribution where the redundancy layers $\eta_k$ follow a power law with exponent $\gamma$ as

$$\hat{r}_u(r) = \frac{r}{1-r}\,\mathrm{Li}_\gamma(1-r) \ .$$

For the special case of $\gamma = 1$, we can use the property $\mathrm{Li}_1(x) = -\ln(1-x)$, and simplify unique recall as

$$\hat{r}_u(r) = -\frac{r\ln r}{1-r} \ .$$

### C.4 Comparing the power law tails

All three power laws show the typical power law straight line in the loglog redundancy frequency plot for their tails (Fig. 4a). The coefficients can be calculated from each other as follows. We first calculate $\gamma = \gamma(\beta)$. From the binomial theorem, we can expand

$$(k+1)^{-\gamma} = k^{-\gamma} - \gamma k^{-\gamma-1} + \frac{\gamma(\gamma-1)}{2}k^{-\gamma-2} - \cdots \ .$$

Applying this formula to Eq. 12, we get

$$\begin{aligned}
\alpha_k &= k^{-1} - (k+1)^{-1} \\
&= \gamma k^{-\gamma-1} - \frac{\gamma(\gamma-1)}{2}k^{-\gamma-2} + \cdots \\
&= \mathcal{O}\left(k^{-(\gamma+1)}\right) \ .
\end{aligned}$$

Hence, we can calculate $\beta$ from $\gamma$ by

$$\beta = \gamma + 1 \ .$$

We next calculate $\delta = \delta(\beta)$. Again using the binomial theorem, this time to $(2k \pm 1)^{-\frac{1}{\delta}}$, we get

$$\begin{aligned}(2k \pm 1)^{-\frac{1}{\delta}} &= \sum_{x=0}^{\infty} \binom{-\frac{1}{\delta}}{x} (2k)^{-\frac{1}{\delta}-x} (\pm 1)^x \\ &= (2k)^{-\frac{1}{\delta}} \pm \left(-\frac{1}{\delta}\right) (2k)^{-\frac{1}{\delta}-1} + \frac{\left(-\frac{1}{\delta}\right)\left(-\frac{1}{\delta}-1\right)}{2} (2k)^{-\frac{1}{\delta}-2} \pm \cdots \ .\end{aligned}$$

Applying this formula to Eq. 11, we get

$$\begin{aligned}\alpha_k &= (2k-1)^{-\frac{1}{\delta}} - (2k+1)^{-\frac{1}{\delta}} \\ &= -2\left(-\frac{1}{\delta}\right) (2k)^{-\frac{1}{\delta}-1} + \cdots \\ &= \frac{2^{-\frac{1}{\delta}}}{\delta} \cdot k^{-\frac{1}{\delta}-1} + \cdots \\ &= \mathcal{O}\left(k^{-\left(\frac{1}{\delta}+1\right)}\right) \ ,\end{aligned}$$

and, hence,

$$\beta = \frac{1}{\delta} + 1 \ .$$

Figure 4e shows the relations between the individual power law exponents.

# D Details Section 5 (evolution of redundancy distributions)

## D.1 Evolution of the uniform distribution

We develop a geometric model of stratified sampling to first deduce the evolution of the uniform distribution $\rho = \text{const}$. We will then generalize this approach in Section D.2 and prove Prop. 2.

We assume the total amount of information to be large. Therefore, our focus can shift from individual pieces of information to fractions of uncountable information where each piece of evidence is infinitesimal. Without loss of generality we set the total amount of unique information to 1. As all information has redundancy $\rho$, the total amount of data is then $\rho$ and we can depict this uniform redundancy distribution as a stack of $\rho$ layers of the same unique information (Fig. 11a). We consider a random sampling process of fraction $r$ in such way, that before sampling, we divide the population into different sub-populations or strata ('strata' means 'layers'), and then take samples from all sub-populations in proportion to their relative sizes. In statistics, this process is known as *stratification*, the process of grouping members of the population into relatively homogeneous and MECE (mutually exclusive, collectively exhaustive) subgroups and then proportional allocation of sample sizes to the subgroups [24]. As, on average, the fractions sampled in a random draw are the same across all strata and the total population, stratification does not change the expected outcome of a sampling process. In our case, where we consider the limit case of very large data sets with $a \to \infty$ data points, this step of subdividing populations and then sampling in proportion to their sizes can be repeated arbitrarily often and does not change the expected outcome of the overall sampling process.

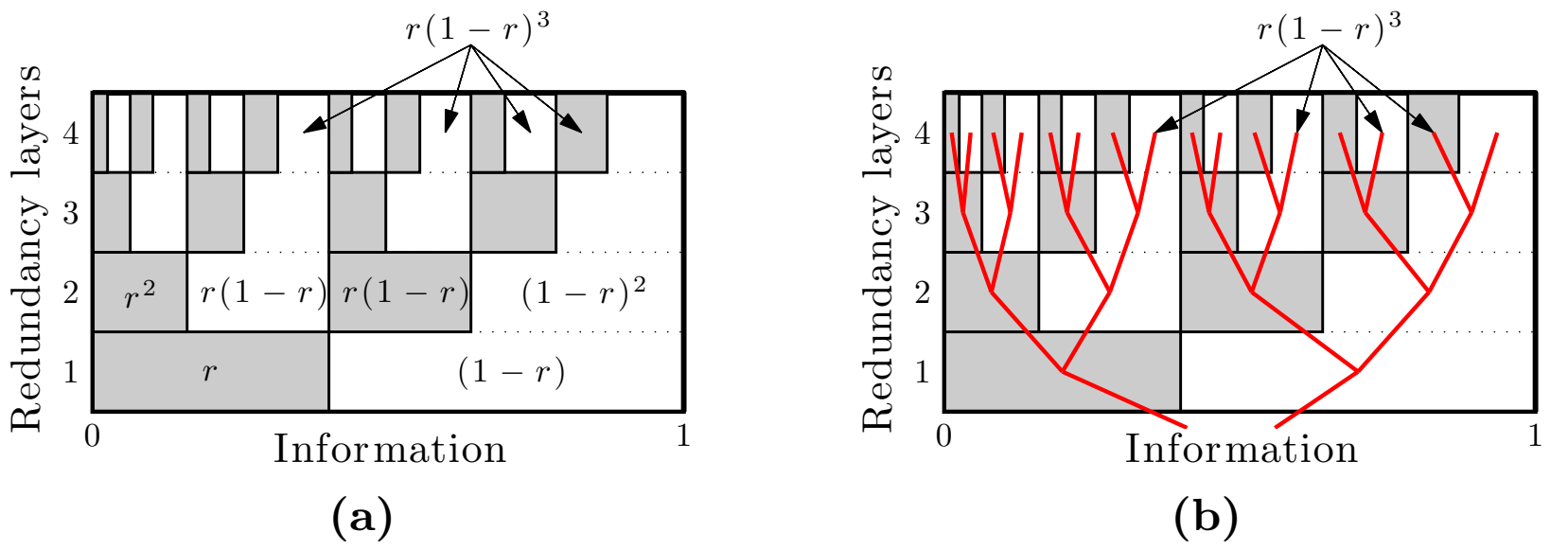

**Fig. 11: (a): Geometric interpretation of sampling from a normalized uniform redundancy distribution: Sampling happens from layer 1 to layer $\rho$, one after the other. In each layer, existing divisions are further divided into two fractions of size proportional to $r$ (grey) and $1-r$ (white). (b): The process of building the divisions from one layer to the next can be compared to a upside-down tree where at each node going left happens with probability $r$ and right with probability $1-r$.**

We start bottom up from layer 1 to layer $\rho$, and at each layer further divide all existing divisions from previous layers into two parts: one of relative size $r$ from which we sample (grey) and one of relative size $1-r$ from which we do not sample (white). At the first layer, we divide into two strata: a fraction $r$ of sampled information and a fraction $1-r$ of unsampled information. In the second layer, we first divide the total amount of information into the same two groups of information already seen in the first layer of size $r$ and a second group with information not yet seen of size $1-r$. Then we choose samples from both subgroups of proportion $r$ of their sizes, thus getting one fraction of size $r^2$ of twice seen information, two strata with size $r(1-r)$ of once seen information, and one strata with size $(1-r)^2$ of not yet sampled information in either layer. Iteratively repeating this process, we have $2^k$ divisions of the total amount of information at any layer $k$. The formation of the divisions in the highest layer $\rho$ can be imagined by the growth of a tree where each division is connected to the division in the previous layer from which it originated (Fig. 11b). The size of each division depends on the number of times it was created by choosing the sample option with proportion $r$ or the not-sample option with size $1-r$. As an example, the arrows in Fig. 11b point to those 4 divisions in the fourth layer which represent one time sampling and three times non-sampling out and which are, therefore, of size $r(1-r)^3$. More general, the size of the divisions in layer $\rho$ representing $k$ times sampling and $\rho-k$ times non sampling is $r^k(1-r)^{\rho-k}$. The number of such divisions is equal to the number of ways that $k$ objects

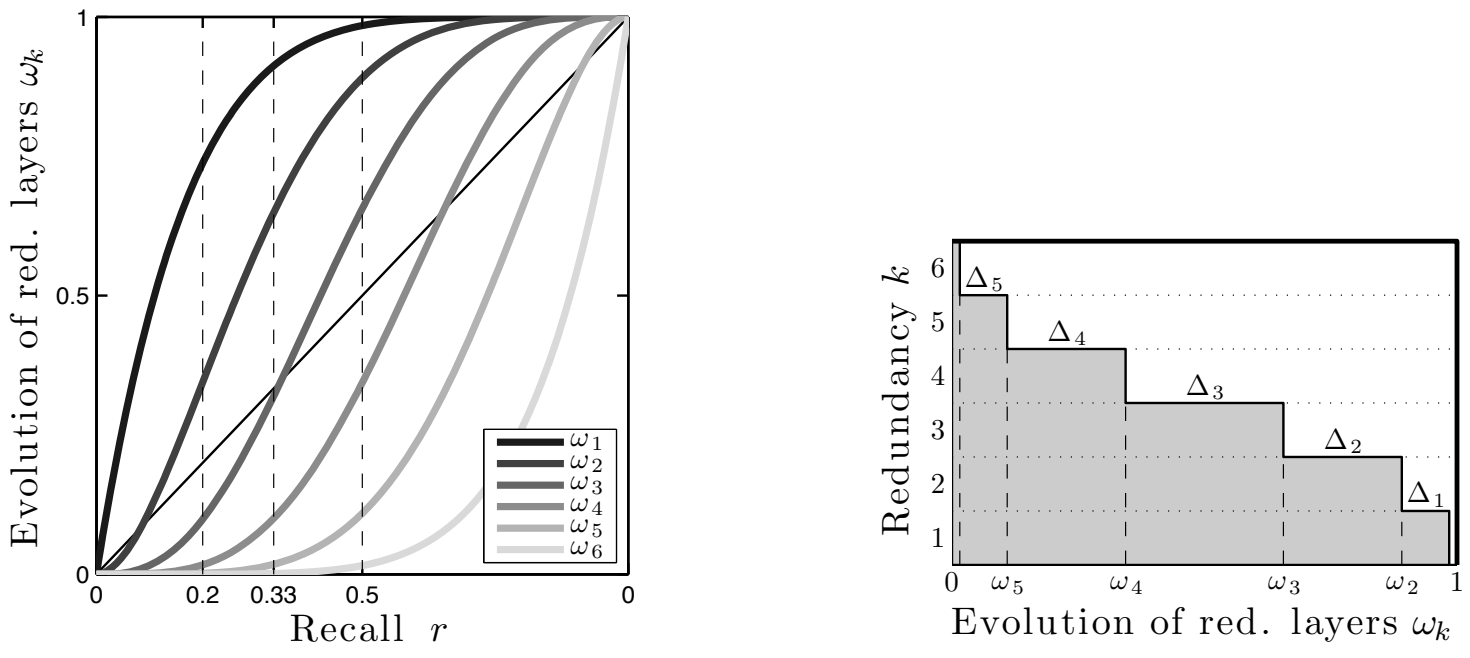

**(a)** **(b)**

**Fig. 12: Evolution of the redundancy layers $\omega_k$, first as function of $r$ (a) and then in the normalized redundancy plot for $r = 0.5$ (b).**

can be chosen from among $\rho$ objects, regardless of order, which is equal to the binomial coefficient $\binom{\rho}{k}$. Multiplying these number, we get as result that the expected fraction of information that appears with redundancy $k$ in a sample from a uniform redundancy distribution with redundancy $\rho$ is equal to the *binomial distribution* $\binom{\rho}{k} r^k (1-r)^{\rho-k}$, which is the probability of getting exactly $k$ successes in $\rho$ independent yes/no experiments, each of which yields success with probability $r$:

$$\varDelta_k(k, \rho, r) = \binom{\rho}{k} r^k (1-r)^{\rho-k} \ .$$

Note that $\varDelta_k(k, \rho, r)$ is actually defined for all $\rho \in \mathbb{N}_1$, $k \in \mathbb{N}_0$ and $r \in [0, 1]$:

$$\varDelta_k(k, \rho, r) = \begin{cases} \binom{\rho}{k} r^k (1-r)^{\rho-k} & \text{if } \ 0 < k < \rho \\ r^\rho & \text{if } \ 0 < k = \rho \\ (1-r)^\rho & \text{if } \ 0 = k < \rho \\ 0 & \text{if } \ k > \rho \ . \end{cases}$$

The evolution $\omega_k$ can then be simply calculated from $\omega_k = 1 - \sum_{y=0}^{k-1} \varDelta_y$ as

$$\boxed{\hat{\omega}_k(k, \rho, r) = 1 - \sum_{y=0}^{k-1} \binom{\rho}{y} r^y (1-r)^{\rho-y}} \ . \qquad (13)$$

Figure 12 illustrates this result.

### D.2 Proof Prop. 2 (evolution of general distributions)

*Proof.* We again use stratification and divide the overall redundancy distribution into homogenous blocks with constant redundancy $x$ and unique information $\alpha_x$

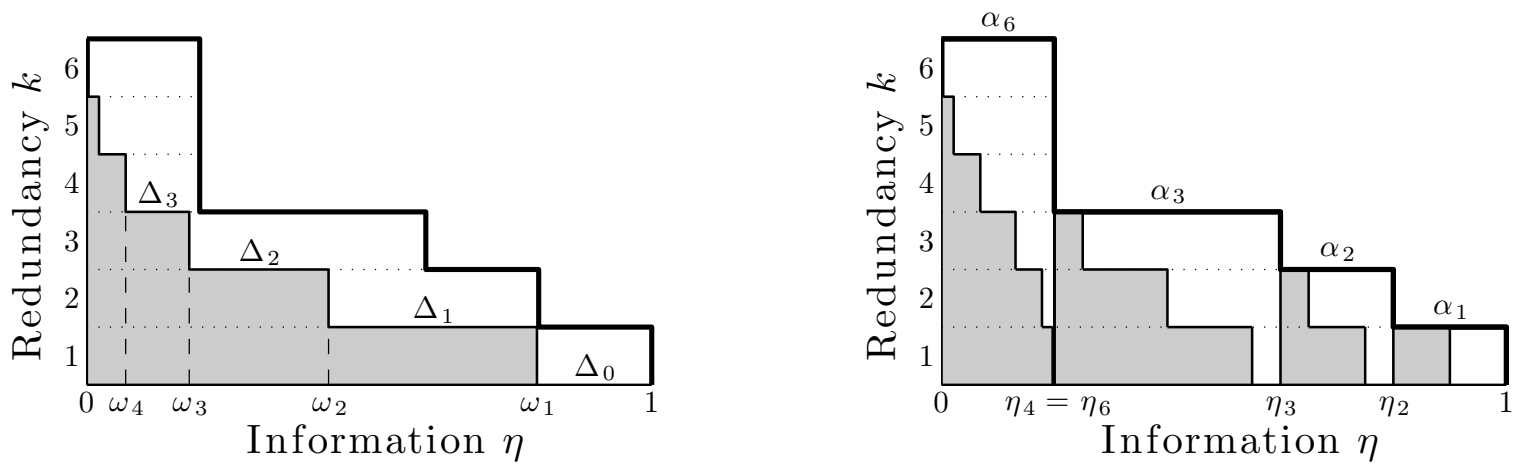

**(a)** **(b)**

**Fig. 13: The evolution $\omega_k$ of the $k$-th redundancy layer of a general redundancy distribution $\boldsymbol{\alpha}$ is the weighted mean of the evolution of this layer in in all uniform distributions.**

before sampling a fraction $r$ from each block in turn (Fig. 13b). The mathematical justification is that, on average, the fractions sampled in a random draw are the same across all sub-populations and the total population [24].

We know that sampling from each block with constant redundancy $x$ follows the previously established relationship of evolution of the uniform (Eq. 13). The amount of information with redundancy $k$ is then the base of the block $\alpha_x$ times $\Delta_k(k, x, r)$. At the same time, the total amount of information $\Delta_k(k, \boldsymbol{\alpha}, r)$ with redundancy $k$ is equal to the sum of $\alpha_x \Delta_k(k, x, r)$ in each block. Hence, the evolution of a general redundancy distribution is the mean of the evolution of of all uniform distributions $x \leq k_{\max}$, weighted by their fractions $\alpha_x$:

$$
\begin{aligned}
\Delta_k(k, \boldsymbol{\alpha}, r) &= \sum_{x=1}^{k_{\max}} \alpha_x \Delta_k(k, x, r) \\
&= \sum_{x=1}^{k_{\max}} \alpha_x \binom{x}{k} r^k (1-r)^{x-k} \ .
\end{aligned}
$$

The evolution of $\boldsymbol{\omega}(k, \boldsymbol{\alpha}, r)$ can again be simply calculated stepwise from $\boldsymbol{\Delta}(k, \boldsymbol{\alpha}, r)$ by $\omega_k = \omega_{k-1} - \Delta_{k-1}$ with $\omega_1 = 1 - \Delta_0$ as

$$
\begin{aligned}
\omega_k &= 1 - \sum_{y=0}^{k-1} \Delta_y \ , \\
&= 1 - \sum_{y=0}^{k-1} \sum_{x=y}^{m} \alpha_x \binom{x}{y} r^y (1-r)^{x-y} \ .
\end{aligned}
$$

Further, k-recall is then given by

$$
r_{uk} = \frac{\omega_k}{\eta_k} \ . \qquad \square
$$

### D.3 Illustration of the limit value of Prop. 2

Figure 14 illustrates that the mathematics developed indeed predicts sampling from large data sets very well. In grey, the individual frames show the *horizontal* evolution of the example normalized redundancy distribution $\boldsymbol{\alpha} = (\frac{1}{5}, \frac{1}{5}, \frac{2}{5}, 0, 0, \frac{1}{5})$ for $r = 0.5$. In red, they show the actual expected *vertical* redundancy distributions as given by Monte Carlo simulations, where the total number of pieces of information $a_u \sim a$ increases from 5 in Fig. 14a to 1000 in Fig. 14d. Note how the horizontal layers of redundancy become visible in the vertical perspective with increasing $a$. These layers become even more apparent in the sample redundancy distribution when taking the median instead of the mean of individual draws in the Monte Carlo simulations (we show here the mean). This observation suggests that the normalized, *horizontal perspective is actually the inherently natural perspective to analyze sampling from large data sets.*

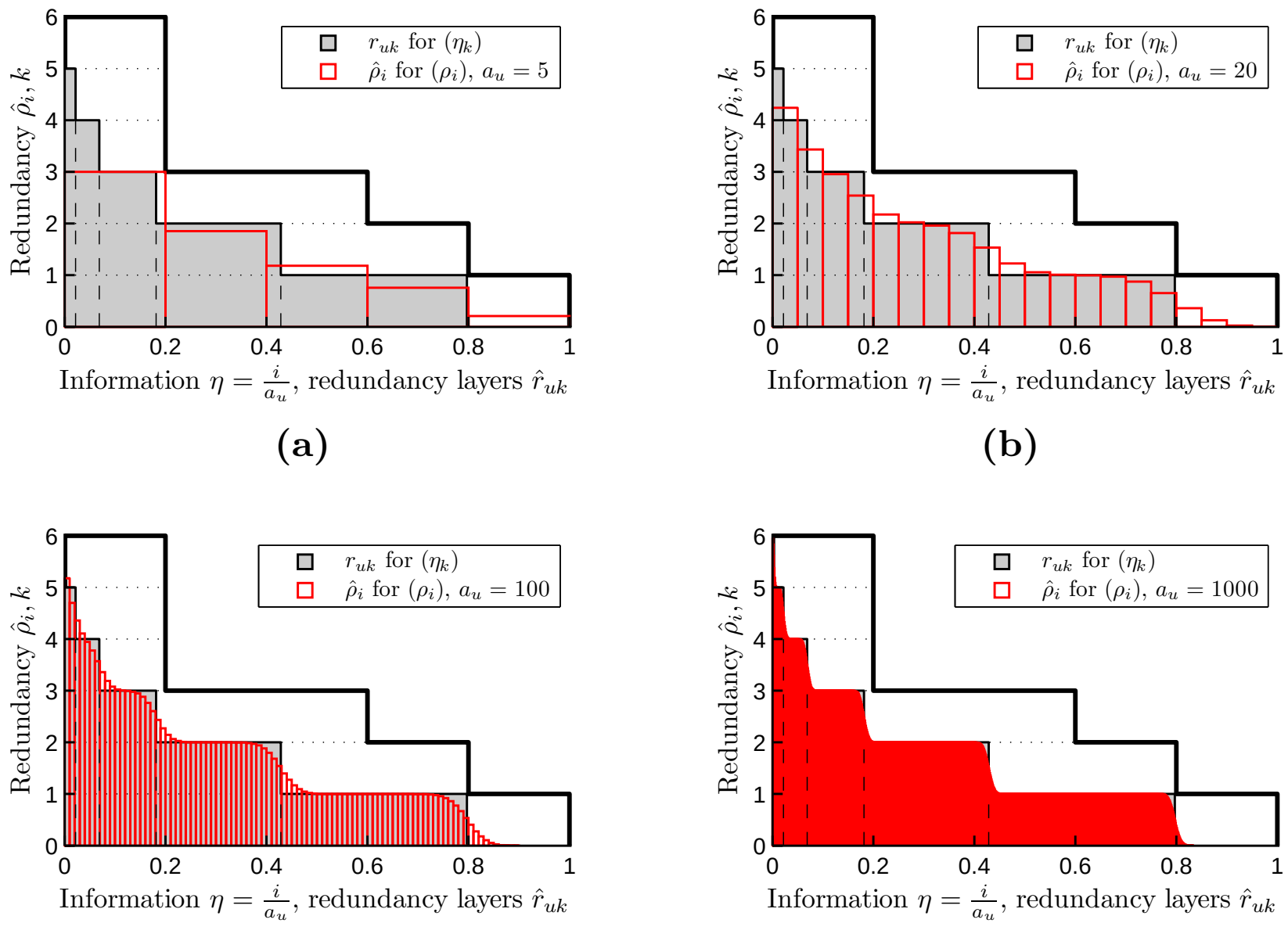

**Fig. 14: Comparing the expected evolution of the redundancy layers $\hat{r}_{uk}$ (grey) with the expected sample redundancy distributions $\hat{\rho}_i$ given by Monte Carlo simulations (red): For large data sets ($a_u \to \infty$), Prop. 2 predicts the sampled distributions increasingly well and the horizontal redundancy layers actually become visible in the vertical sample distributions.**

# E Details Section 6 (evolution of power laws)

## E.1 Proof Lemma 1 (evolution of the invariant)

*Proof.* We prove that the redundancy distribution $\boldsymbol{\alpha}_I = (\alpha_k)$ with

$$\alpha_k = (-1)^{k-1}\binom{\tau}{k}, \text{ and } 0 < \tau \leq 1 ,$$

is invariant under sampling. We show this by proving that the following holds for $\boldsymbol{\alpha}_I$:

$$r_{uk}(k, \boldsymbol{\alpha}, r) = r^{\tau} .$$

The statement then follows from $r_{uk}$ being independent of $k$ and $r_{uk} = r_{u1} = \omega_1$. We make use of the following easy-to-verify identity

$$\binom{i}{k}\frac{\binom{\tau}{i}}{\binom{\tau}{k}} = \binom{\tau - k}{i - k} ,$$

and use $\theta_k = \frac{\Delta_k}{\alpha_k}$ to describe the fraction of information with redundancy equal $k$ in the sample to that with redundancy equal $k$ in the original distribution. Starting from

$$\Delta_k(k, \boldsymbol{\alpha}, r) = \sum_{i=k}^{\infty} \Delta_k(k, i, r)\alpha_i ,$$

we can write $\theta_k = \Delta_k / \alpha_k$ as

$$\theta_k(k, \boldsymbol{\alpha}, r) = \sum_{i=k}^{\infty} \Delta_k(k, i, r)\frac{\alpha_i}{\alpha_k} .$$

Then, for $\boldsymbol{\alpha}_I = (\alpha_k)$

$$\begin{aligned}
\theta_k(k, \boldsymbol{\alpha}_I, r) &= \sum_{i=k}^{\infty} \binom{i}{k} r^k (1-r)^{i-k} \frac{(-1)^{i-1}\binom{\tau}{i}}{(-1)^{k-1}\binom{\tau}{k}} \\
&= \sum_{i=k}^{\infty} r^k (1-r)^{i-k} (-1)^{i-k} \binom{\tau - k}{i - k} \\
&= r^k \sum_{i=k}^{\infty} (r-1)^{i-k} \binom{\tau - k}{i - k} \\
&= r^k \sum_{i=0}^{\infty} (r-1)^{i} \binom{\tau - k}{i} \\
&= r^k \left((r-1) + 1\right)^{\tau - k} \\
&= r^{\tau} .
\end{aligned}$$

Further,

$$\omega_k(k, \boldsymbol{\alpha}_I, r) = \sum_{x=k+1}^{\infty} \Delta_x = \sum_{x=k+1}^{\infty} \theta_x \alpha_x = r^\tau \eta_k \ ,$$

and, hence,

$$r_{uk}(k, \boldsymbol{\alpha}_I, r) = r^\tau \text{ , with } 0 < \tau \leq 1 \ . \qquad \square$$

### E.2 Necessary condition for invariants during sampling

We next argue that the invariant family is the only type of redundancy distribution that remains invariant under sampling. We proceed in two step. We first give an conjectured property that must hold for every distribution that is invariant. As our argumentation does not stand the requirements of a rigorous proof, we call this Conjecture 1. We then show with Corollary 1, if this conjecture is true, then the invariant family is indeed the only redundancy distribution invariant under sampling without replacement.

**Conjecture 1 (Necessary condition for invariants).** *The following is a necessary condition for an invariant of sampling*

$$r_{uk}(k, \boldsymbol{\alpha}, r) = r^\tau \text{ with } 0 < \tau \leq 1 \ . \tag{14}$$

*Intuitive argument for Conjecture 1.* If a function remains invariant during the evolution, then we know that the k-recall $r_{uk}$ is the same for each redundancy layer $k$. Now while the overall recall $r$ grows, the total amount of sampled data has to be accommodated by the "space" formed by the growing layers of redundancy. This space is formed by the dimensionality of the shape of the distribution. While this shape is filled with more and more data, unique recall has to grow according to some function that simulated this filling of the space. Comparing such a shape with a higher dimensional triangle or tetrahedron of higher dimension (Fig. 15a and Fig. 15b), the functions would be the $n$-th square for dimension $n$, which translates in a function that grows according to

$$r_{uk}(\boldsymbol{\alpha}_I, r) = r^{1/n} \ .$$

Since unique recall $r_{u1}$ is concave and always bigger than $r$ except for $r \in \{0, 1\}$, $n$ must be bigger than 1. Hence, the following condition must hold

$$r_{uk}(\boldsymbol{\alpha}_I, r) = r^\tau \text{ , with } 0 < \tau \leq 1 \ . \tag{15}$$

Figure 15c and Fig. 15d show such an example invariant redundancy distribution with $n = 2$ and the resulting unique recall function with $r_u = \sqrt{r}$. The function could therefore be called a function of dimensionality 2.

**Corollary 1 (Invariant family).** *No other distribution than the distributions defined in Lemma 1 is invariant under sampling.*

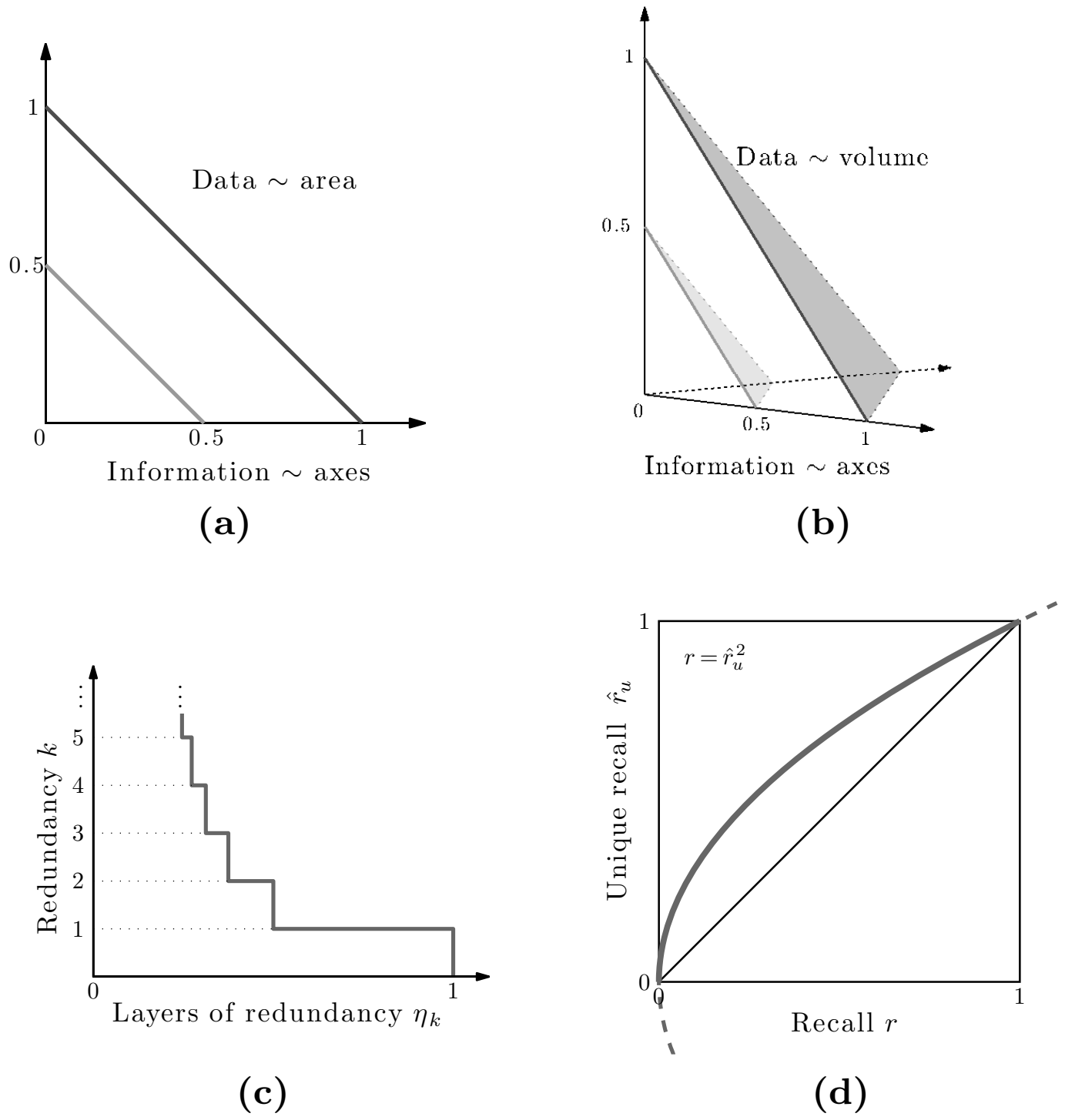

**Fig. 15: Intuition for Corollary 1: When sampling from a smooth and regular $n$-dimensional data space, the only function that can describe a realistic concave sampling success is the $n$-th root, thus giving $\text{information} = \sqrt[n]{\text{data}}$.**

*Proof.* Assume that Eq. 15 holds. We show that the invariant family of Lemma 1 necessarily follows from this conjecture. First, we notice that Eq. 15 also has to hold for $k = 1$ and, hence, we have with $\eta_1 = 1$,

$$r_u(\boldsymbol{\alpha}_I, r) = r^{\tau} \ .$$

Calculating the derivatives of $r_u(r)$, we get

$$\begin{aligned} r_u(r) &= r^{\tau} \\ r'_u(r) &= \tau\, r^{\tau-1} \\ r''_u(r) &= \tau(\tau-1)\, r^{\tau-2} \\ &\vdots \\ r_u^{(k)}(r) &= \tau^{\underline{k}}\, r^{\tau-k} \ , \end{aligned}$$

and for the interesting point $r = 1$,

$$r_u^{(k)}(r)\Big|_{r=1} = \tau^{\underline{k}} \ .$$

At the same time, we get from Prop. 1:

$$\begin{aligned}
\hat{r}_u(r) &= 1 - \sum_{k=1}^{m} \alpha_k \, (1-r)^k \\
\hat{r}'_u(r) &= \sum_{k=1}^{m} k\alpha_k \, (1-r)^{k-1} \\
\hat{r}''_u(r) &= -\sum_{k=2}^{m} k(k-1)\alpha_k \, (1-r)^{k-2} \\
&\vdots \\
\hat{r}^{(n)}_u(r) &= (-1)^{n-1} \sum_{k=n}^{m} k^{\underline{n}} \, \alpha_k \, (1-r)^{k-n} \quad .
\end{aligned} \tag{16}$$

The term $k^{\underline{n}}$ in the last equation stands for the falling factorial powers (or short, "falling factorial" or "$k$ to the $n$ falling") $k^{\underline{n}} = k(k-1)\cdots(k-n+1)$ [17, p.47]. For the end point $r = 1$, or actually taking the limit value for $r \to 1$ of Eq. 16, we get that, in the limit, all terms in the sum disappear except for $k = n$:

$$\hat{r}^{(n)}_u(r) \bigg|_{r=1} = \lim_{r \to 1} \hat{r}^{(n)}_u(r) = (-1)^{n-1} n! \, \alpha_n \ ,$$

and we can express $\alpha_k$, the fraction of information with redundancy $k$, as simple function of the $k^{\text{th}}$ derivative of $r_u$:

$$\alpha_k = (-1)^{k-1} \frac{\hat{r}^{(k)}_u(1)}{k!} \quad .$$

From that, we can now calculate $\boldsymbol{\alpha}_I$ as

$$\begin{aligned}
\alpha_k &= (-1)^{k-1} \frac{\hat{r}^{(k)}_u(1)}{k!} \\
&= (-1)^{k-1} \frac{\tau^{\underline{k}}}{k!} \ ,
\end{aligned}$$

where the last equation can be written as

$$\alpha_k = (-1)^{k-1} \binom{\tau}{k} \ ,$$

since the binomial coefficient is defined for all $\tau \in \mathbb{R}$ [20, p.51].

### E.3 Proof Lemma 2 (sampling from complete power laws)

The statement of the tail remaining invariant during evolution is equivalent to

$$\lim_{i,j \to \infty} \frac{r_{ui}}{r_{uj}} = 1 \ ,$$

and, hence $r_{uk}$ being independent of $k$ for large $k$. This is what we prove in two steps. First we have to prove the limit value in the following Lemma 4, then use this limit value to prove Lemma 2.

**Lemma 4 (Convergence of $\Theta(k,i,\beta)$).**

$$\lim_{k\to\infty} \underbrace{\frac{(i-\beta)^{\underline{i-k}}}{i^{\underline{i-k}}} \left(\frac{k}{i}\right)^{-\beta}}_{\Theta(k,i,\beta)} = 1 \text{ , for } i \geq k, \beta > 1 \text{ ,} \tag{17}$$

*Proof.*

$$\begin{aligned}
\Theta(k,i,\beta) &= \left(\frac{k}{i}\right)^{-\beta} \frac{(i-\beta)^{\underline{i-k}}}{i^{\underline{i-k}}} \\
&= \left(\frac{k}{i}\right)^{-\beta} \frac{(k-\beta+1)^{\overline{i-k}}}{(k+1)^{\overline{i-k}}} && |\text{transform to rising factorial} \\
&= \left(\frac{k}{i}\right)^{-\beta} \frac{(k-\beta)^{\overline{i-k+1}}}{(k)^{\overline{i-k+1}}} \frac{k}{k-\beta} && |\beta \neq k \\
&= \left(\frac{k}{i}\right)^{-\beta} \frac{(1-\beta)^{\overline{i}}}{(1)^{\overline{i}}} \frac{(1)^{\overline{k}}}{(1-\beta)^{\overline{k}}} && |\beta \notin \{1,\ldots,k\} \\
&= \underbrace{\frac{k^{-\beta}k!}{(1-\beta)^{\overline{k}}}}_{\Omega(k,\beta)} \underbrace{\frac{(1-\beta)^{\overline{i}}}{i^{-\beta}i!}}_{\Psi(i,\beta)}
\end{aligned}$$

Note that $\Psi(k,\beta) = \Omega(k,\beta)^{-1}$. Hence, to show $\lim_{k\to\infty} \Theta(k,i,\beta) = 1$ for $i \geq k$, it suffices to show (i) that $\Omega(k,\beta)$ monotonically increases in $k$ above a certain $k_0$, and (ii) that $\Omega(k,\beta)$ converges for $k \to \infty$.

Monotonicity follows from direct calculation:

$$\begin{aligned}
\Omega(k+1,\beta) &> \Omega(k,\beta) \\
\frac{(k+1)^{-\beta}(k+1)!}{(1-\beta)^{\overline{k}}(k+1-\beta)} &> \frac{k^{-\beta}k!}{(1-\beta)^{\overline{k}}} \\
\frac{k+1}{k+1-\beta} &> \left(\frac{k+1}{k}\right)^{\beta} \\
\frac{j}{j-\beta} &> \left(\frac{j}{j-1}\right)^{\beta} && |j := k-1 \geq 2 \\
(j-1)^{\beta} &> (j-\beta)j^{\beta-1} \\
j^{\beta} - \binom{\beta}{1} j^{\beta-1} + \binom{\beta}{2} j^{\beta-2} - \ldots &> j^{\beta} - \beta j^{\beta-1} && |\text{binomial theorem} \\
\binom{\beta}{2} j^{\beta-2} - \ldots &> 0 \text{ ,}
\end{aligned}$$

which is true for $j > \beta - 1$, and, hence $k > k_0 = \max[\beta, 2]$.

Convergence for $\beta \notin \mathbb{N}$ follows from

$$\lim_{n\to\infty} \Omega(n,\beta) = \Gamma(1-\beta) \text{ ,}$$

which follows from Euler's formula for the Gamma function [1, Eq. 6.1.2]

$$\Gamma(z) = \lim_{n\to\infty} \frac{n^z n!}{(z)^{\overline{n+1}}} \ .$$

or

$$\Gamma(1-\beta) = \lim_{n\to\infty} \frac{n^{1-\beta} n!}{(1-\beta)^{\overline{n+1}}} \ .$$

$$\begin{aligned}
\lim_{n\to\infty} \Omega(n,\beta) &= \lim_{n\to\infty} \frac{n^{-\beta} n!}{(1-\beta)^{\overline{n}}} \\
&= \lim_{n\to\infty} \frac{n^{1-\beta} n!}{(1-\beta)^{\overline{n+1}}} \frac{n+1-\beta}{n} \\
&= \Gamma(1-\beta) \lim_{n\to\infty} \frac{n+1-\beta}{n} \\
&= \Gamma(1-\beta) \ .
\end{aligned}$$

In the above derivation, we had to state $\beta \notin \mathbb{N}$ due to the otherwise undefined value of $\frac{0}{0}$. To include all $\beta > 1$, we can note

$$\Theta(k,i,\beta) = \underbrace{\frac{(z-\beta)^{\overline{i-z+1}}}{i^{-\beta} i!}}_{\Psi(i,\beta,z)} \underbrace{\frac{k^{-\beta} k!}{(z-\beta)^{\overline{k-z+1}}}}_{\Omega(k,\beta,z)} \ , \tag{18}$$

where $z \in \mathbb{N}$ is chosen so $i \geq k \geq z > \beta > 1$.

*Proof (Lemma 2).* From Fig. 4, we know that the three power laws have the same tail distributions. Without loss of generality, we consider here power laws in the redundancy frequency plot $\boldsymbol{\alpha}_C$ with

$$\alpha_k = \frac{k^{-\beta}}{\zeta(\beta)} \ .$$

Starting from

$$\Delta_k(k,\boldsymbol{\alpha},r) = \sum_{i=k}^{\infty} \Delta_k(k,i,r)\alpha_i \ ,$$

we can write $\theta_k = \Delta_k/\alpha_k$ as

$$\theta_k(k,\boldsymbol{\alpha},r) = \sum_{i=k}^{\infty} \Delta_k(k,i,r)\frac{\alpha_i}{\alpha_k} \ ,$$

and, for $\boldsymbol{\alpha}_C = (\alpha_k)$

$$\theta_k(k,\boldsymbol{\alpha}_C,r) = \sum_{i=k}^{\infty} \binom{i}{k} r^k (1-r)^{i-k} \left(\frac{i}{k}\right)^{-\beta} \ .$$

From Eq. 17, we know

$$\begin{aligned}
\lim_{k\to\infty}\binom{i}{k}\left(\frac{i}{k}\right)^{-\beta} &= \binom{i}{k}\frac{(i-\beta)^{\underline{i-k}}}{i^{\underline{i-k}}} \\
&= \binom{i}{k}\frac{(i-\beta)\cdots(k-\beta+1)}{i^{\underline{i-k}}} \\
&= \binom{i}{k}\frac{(\beta-k-1)\cdots(\beta-i)(-1)^{(i-k)}}{i^{\underline{i-k}}} \\
&= \binom{i}{k}\frac{(\beta-k-1)^{\underline{i-k}}}{i^{\underline{i-k}}}(-1)^{(i-k)} \\
&= \frac{i!}{(i-k)!k!}\frac{(\beta-k-1)^{\underline{i-k}}}{i^{\underline{i-k}}}(-1)^{(i-k)} \\
&= \frac{(\beta-k-1)^{\underline{i-k}}}{(i-k)!}(-1)^{(i-k)} \\
&= \binom{\beta-k-1}{i-k}(-1)^{(i-k)} \ .
\end{aligned}$$

We can now write

$$\begin{aligned}
\lim_{k\to\infty}\theta_k(k,\boldsymbol{\alpha}_C,r) &= \sum_{i=k}^{\infty} r^k(1-r)^{i-k}(-1)^{i-k}\binom{\beta-k-1}{i-k} \\
&= r^k\sum_{i=k}^{\infty}(r-1)^{i-k}\binom{\beta-k-1}{i-k} \\
&= r^k\sum_{i=0}^{\infty}(r-1)^{i}\binom{\beta-k-1}{i} \\
&= r^k\left((r-1)+1\right)^{\beta-k-1} \\
&= r^{\beta-1} \ .
\end{aligned}$$

Then,

$$\begin{aligned}
\lim_{k\to\infty}\omega_k(k,\boldsymbol{\alpha}_C,r) &= \lim_{k\to\infty}\sum_{x=k+1}^{\infty}\Delta_x \\
&= \lim_{k\to\infty}\sum_{x=k+1}^{\infty}\theta_x\alpha_x \\
&= r^{\beta-1}\eta_k \ ,
\end{aligned}$$

and, hence,

$$\lim_{k\to\infty} r_{uk}(k,\boldsymbol{\alpha}_C,r) = r^{\beta-1} \ ,$$

which is independent of $k$. □

Figure 16 illustrates Lemma 2 for 3 power law coefficients.

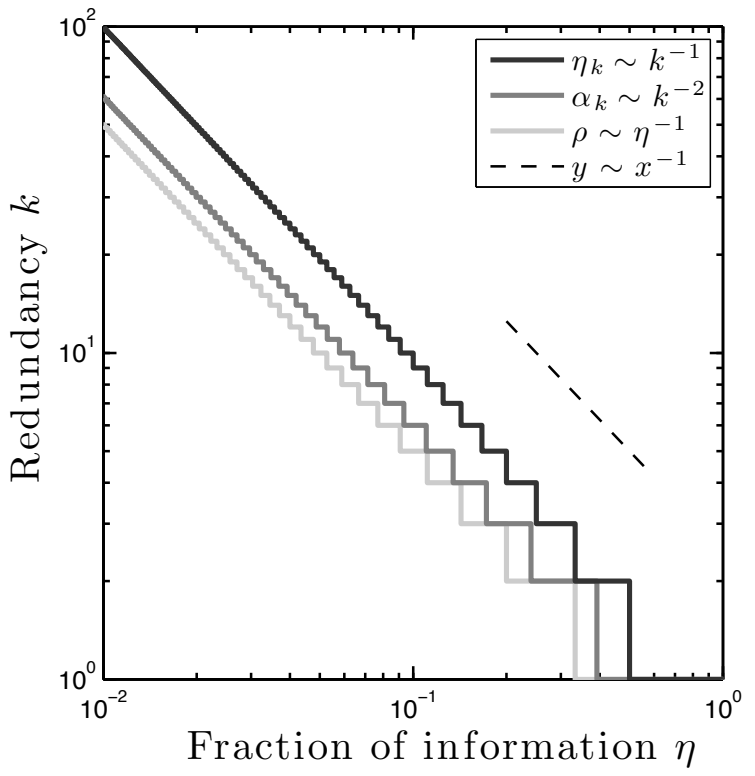

**(a)** $\gamma = 1$ ($\beta = 2$, $\delta = 1$)

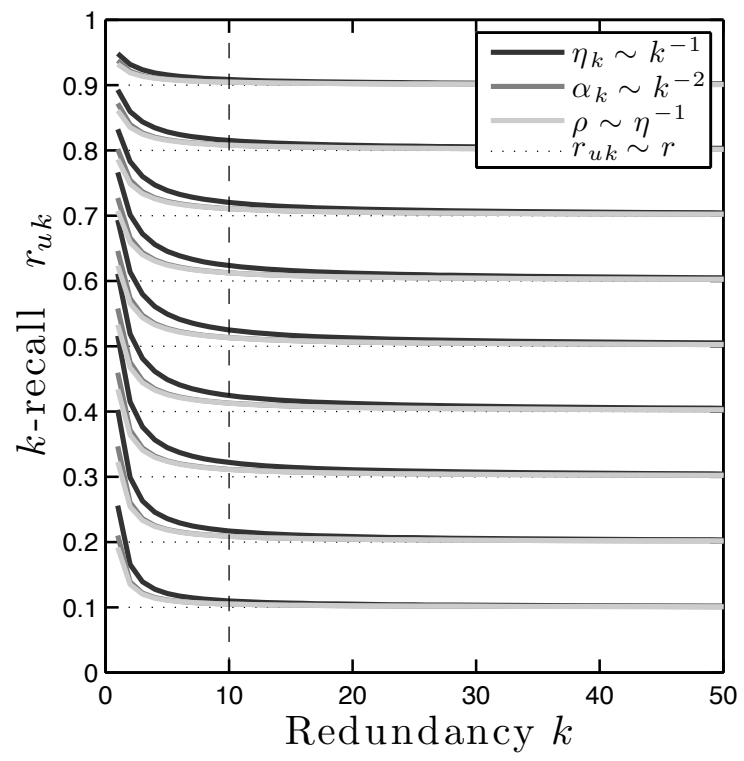

**(b)** $\gamma = 1$ ($\beta = 2$, $\delta = 1$)

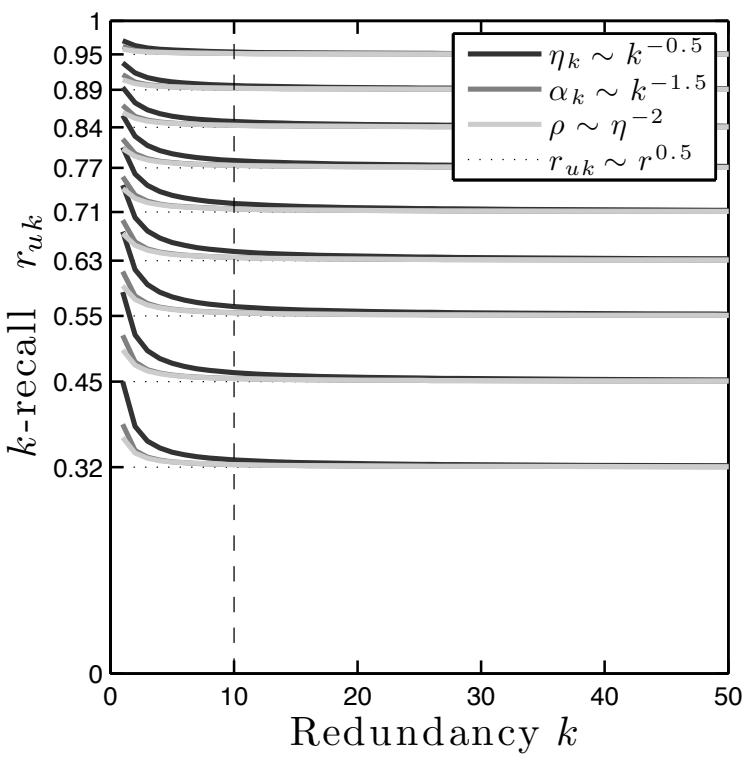

**(c)** $\gamma = 0.5$ ($\beta = 1.5$, $\delta = 2$)

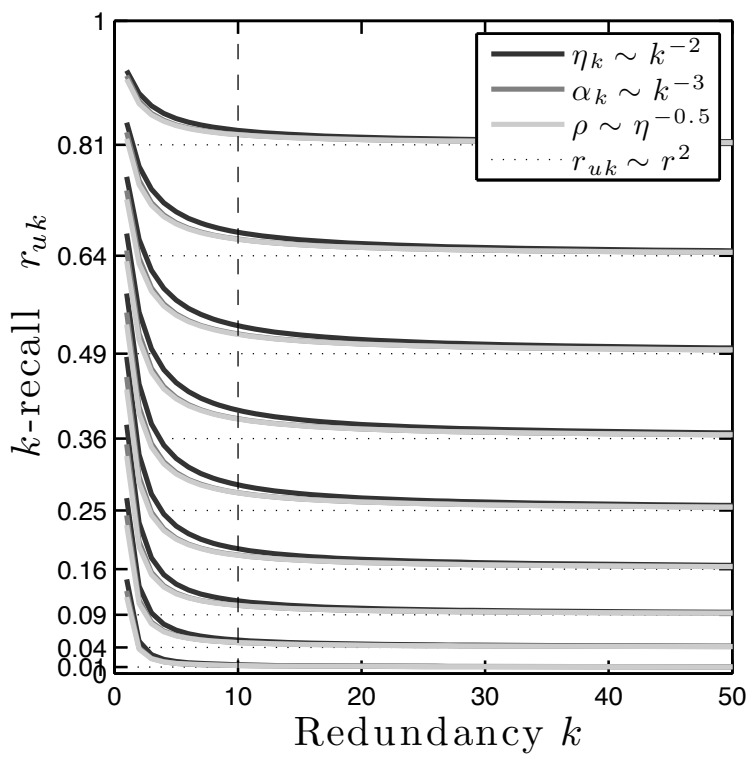

**(d)** $\gamma = 2$ ($\beta = 3$, $\delta = 0.5$)

**Fig. 16: Sampling from any completely developed power law leads to other power laws since k-recall $r_{uk}$ is independent of $k$ above a certain threshold.**

## F Proof Theorem 1 (sampling from truncated power laws)

*Proof.* Theorem 1 follows readily from Lemma 2. Assume $\boldsymbol{\alpha}_T$ is a truncated power law with maximum redundancy $k_{\max}$. (1) for $k > k_{\max}$ $\Delta_k = 0$ by definition. (2) For $k \leq k_{\max}$ we know

$$\begin{aligned} \Delta_k(k, \boldsymbol{\alpha}, r) &= \sum_{i=k}^{k_{\max}} \Delta_k(k, i, r)\alpha_i \\ &= \sum_{i=k}^{\infty} \Delta_k(k, i, r)\alpha_i - \sum_{i=k_{\max}}^{\infty} \Delta_k(k, i, r)\alpha_i \ , \end{aligned}$$

(2a) For $k \ll k_{\max}$, the second summand is small as compared to the first one and we get

$$\lim_{k \ll k_{\max}} \Delta_k(k, \boldsymbol{\alpha}_T, r) = \Delta_k(k, \boldsymbol{\alpha}_C, r)$$

from which follows that at the lower side, the sample distribution from a truncated power law behaves the same as from a completely developed power law.

(2b) For $k \to k_{\max}$ the second term becomes increasingly dominant and $\Delta_k$ and, hence, $r_{uk}$ too become smaller. If now $k_{\max}$ is sufficiently enough (large data sets), the observed distribution must have an observed power law distribution. □

Figure 17, together with Figure 5c and Fig. 5d illustrate Theorem 1.

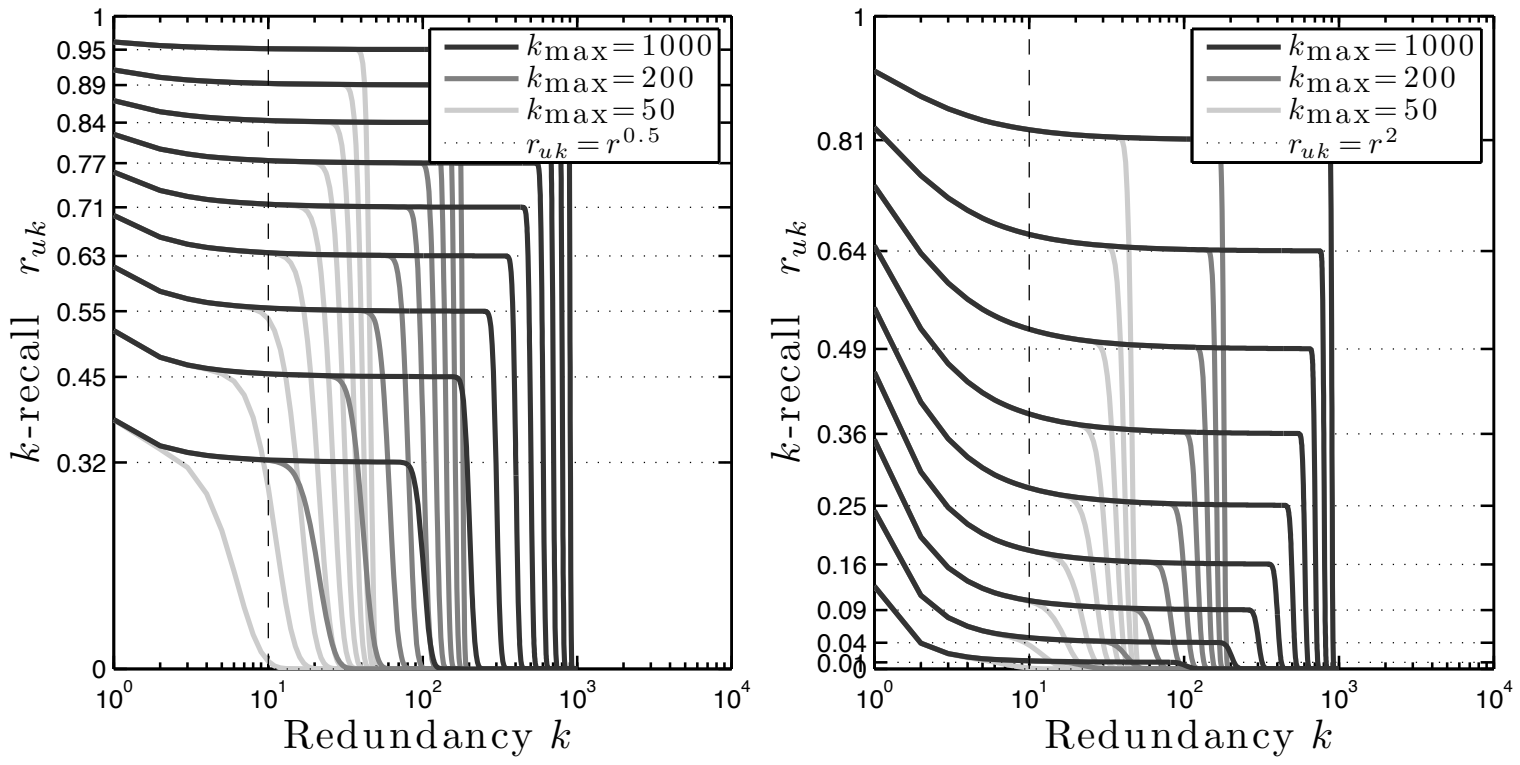

**(a)** $\gamma = 1$ **(b)** $\gamma = 1$

**Fig. 17: Theorem 1: Sampling from truncated power law distributions leads to power law distributions with the tail "breaking in" for increasingly lower recalls. However, the core of the power law still shows $r_{uk} \approx r^{\gamma}$, and the sample distribution thus is a power law. The larger the data set, the better the approximation.**